\documentclass{article} %
\usepackage{iclr2027_conference,times}

\usepackage{amsmath,amsfonts,bm}

\def\eqref#1{equation~\ref{#1}}

\def\1{\bm{1}}

\DeclareMathAlphabet{\mathsfit}{\encodingdefault}{\sfdefault}{m}{sl}
\SetMathAlphabet{\mathsfit}{bold}{\encodingdefault}{\sfdefault}{bx}{n}

\definecolor{mydarkred}{rgb}{0.6,0,0}
\definecolor{myblue}{HTML}{268BD2}
\usepackage[colorlinks, linkcolor=mydarkred, citecolor=myblue]{hyperref}
\usepackage{etoc}
\usepackage{url}
\usepackage{enumitem}
\usepackage{algorithm}
\usepackage{algpseudocode}
\usepackage{wrapfig}
\usepackage{booktabs}
\usepackage{graphicx}
\usepackage{pifont}
\newcommand{\yes}{\textcolor{green!55!black}{\ding{51}}}
\newcommand{\no}{\textcolor{red!80!black}{\ding{55}}}

\title{Recall Begets Recall: Recurrent Memory Ranking for Personalized LLMs}

\author{%
Pengyue Jia$^{1*}$, Shida Yan$^{1*}$, Shunyu Liu$^{2}$, Yingyi Zhang$^{1}$, Leszek Rutkowski$^{3}$,\\
\textbf{Chenyang Lyu$^{4}$, Dacheng Tao$^{2}$} \\
\textnormal{$^{1}$City University of Hong Kong \quad $^{2}$Nanyang Technological University} \\
\textnormal{$^{3}$AGH University of Science and Technology \quad $^{4}$Alibaba Group} \\
}

\iclrfinalcopy %
\begin{document}

\maketitle
\lhead{Preprint.} %
{\renewcommand{\thefootnote}{}\footnotetext{$^{*}$Equal contribution.}}

\begin{abstract}
Memory ranking is the stage in personalized Large Language Models (LLMs) that decides which of the candidate memories returned by a retriever enter the limited context, so that the answer is grounded in the right parts of the user's history.
Existing memory rankers operate in a single pass, scoring each candidate once against the raw query and cutting the ranked list at a fixed top-$K$.
The single pass misses memories that the answer needs but that look unrelated to the query on their own, and the fixed cut ignores how many memories the query needs.
To address both problems, we propose \textbf{ReMind}, a memory ranker that runs as a \textit{recurrent} and \textit{self-terminating} loop, following the cognitive theory of memory search: each recalled item joins the cue for the next recall (compound cueing), and recall stops once enough has been collected for the task (metamemory monitoring).
For compound cueing, ReMind adds the top-ranked memory to the query context at each step and reranks the remaining candidates under the expanded cue, 
so that each recall makes the cue more specific and later steps look for what is still missing rather than for what merely resembles the query.
For metamemory monitoring, a sufficiency predictor estimates at each step whether the collected memories are enough to answer the query and stops the loop once they are, so that the number of selected memories adapts to each query.
Experiments on public benchmarks show that ReMind achieves the best results compared to state-of-the-art memory rankers.
\end{abstract}

\section{Introduction}
\label{sec:introduction}

LLMs are increasingly deployed as long-term dialogue agents that serve the same user across many sessions, such as personal assistants~\citep{zhang2025survey} and customer service agents~\citep{tan2025prospect, chhikara2025mem0}.
In these applications, a helpful response often depends on information the user provided in an earlier session, such as a stated preference~\citep{xu2022beyond,zhao2025llms} or a past decision~\citep{tan2025prospect}.
To keep this information available, recent work pairs the LLM with a memory system that stores the interaction history as a collection of memories~\citep{chhikara2025mem0}. At answer time, memories relevant to the current query are retrieved and placed into the model context \citep{ xu2026mem,pan2025secom}.
Considering the token consumption and the context limit of LLMs, only a small set of memories can be placed in the context~\citep{jin2025long}, so the quality of the answer depends directly on which memories are selected~\citep{liu2024lost,shi2023large}.
This selection is far from easy: as the history grows across sessions, the memories of a user become numerous and noisy, and in some scenarios the accumulated history reaches over a million tokens \citep{maharana2024evaluating, wu2024longmemeval,tavakoli2025beyond}.
Finding the few memories that the query needs from such a history is thus a challenging problem~\citep{hu2026evaluating}.

\begin{figure}[t]
    \centering
    \includegraphics[width=\linewidth]{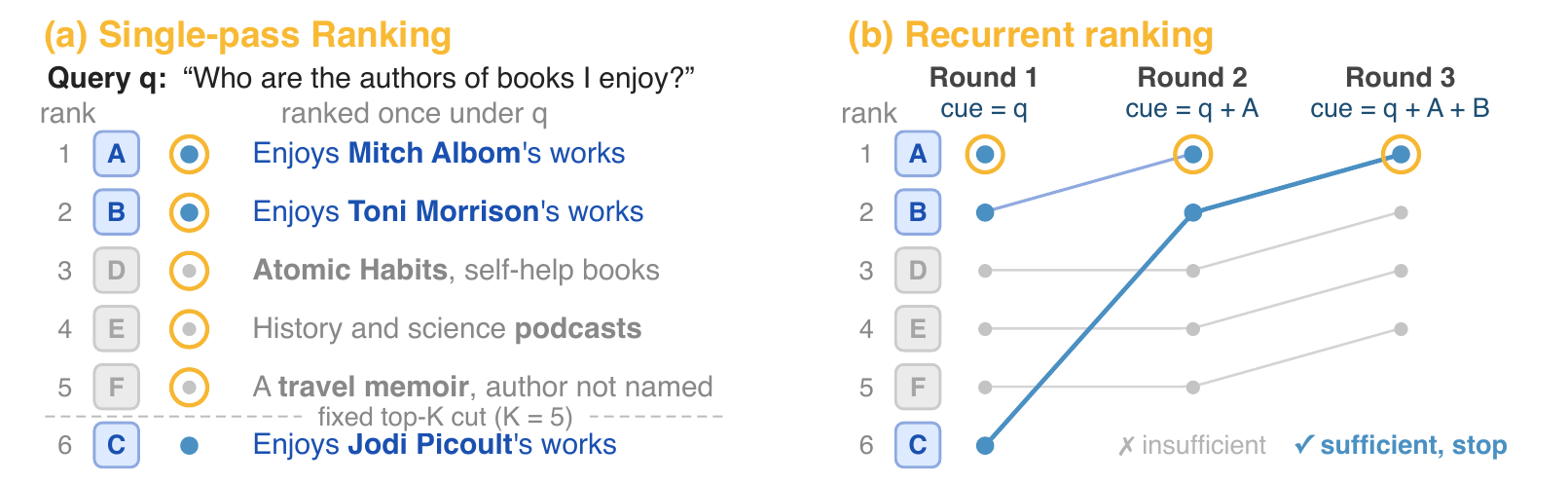}
    \caption{Comparison of single-pass and recurrent ranking on an example from PersonaBench.}
    \label{fig:remind-motivation}
\end{figure}

Existing memory systems commonly adopt a retrieval-then-ranking pipeline to find the related memories~\citep{tan2025prospect,ma2023finetuning}.
Given the candidates retrieved by a lightweight retriever, ranking is the stage that makes the final selection~\citep{nogueira2019passage}: it models the fine-grained relation between the query and each candidate, and decides which memories enter the context.
Generic rankers~\citep{zhuang2023rankt5} score each candidate by its semantic similarity to the query; however, a semantically similar memory does not always contain the information the answer needs~\citep{cuconasu2024power,zhang2024large}: for the query in Figure~\ref{fig:remind-motivation}, the memories about a self-help book and a travel memoir are close to the query in topic, yet neither of them helps answer the question.
Therefore, recent work trains rankers on memory data so that they fit the memory scenario better \citep{li2026memreranker, li2026query, pan2026convmemory, tan2025prospect}.
These trained memory rankers mainly share the same template: each candidate is scored once against the raw query in a single pass, and the top-$K$ memories are incorporated into the context for response.

This template has two problems, and both come from the same root: the ranker scores every candidate with the query alone and never looks at what it has already selected.
First, the relevance of each memory is judged in isolation.
The memories that a personalized query needs are often related to each other more than to the query itself, and a memory may become clearly relevant only after another one has been selected; a ranker that sees only the query has no way to reach such memories, as with the third author in Figure~\ref{fig:remind-motivation} (a).
Second, the ranker has no way to tell when it has collected enough, so the cut has to be fixed in advance.
A fixed top-$K$ is too large for a query that needs one memory, filling the context with distracting memories and extra tokens, and too small for a query that needs many, leaving the answer without evidence.
In contrast, cognitive models of memory search describe human recall as a recurrent and self-terminating process, in which each recalled item joins the cue for the next recall (compound cueing)~\citep{raaijmakers1981search,howard2002distributed,polyn2009context} and the search stops once enough has been recalled for the task (metamemory monitoring) \citep{ nelson1990metamemory}.
This raises the central question of this work: \textit{can we build a memory ranker that works the same way, conditioning each ranking step on the memories already selected and stopping once they are sufficient?}

To answer this question, we propose \textbf{ReMind}, a memory ranker that performs ranking as a recurrent and self-terminating loop, as shown in Figure~\ref{fig:remind-motivation} (b).
To realize compound cueing, ReMind promotes the top-ranked memory into the query context at each step and reranks the remaining candidates under the expanded cue, so that memories selected early serve as extra cues for memories that the raw query alone cannot match. Each step therefore looks for what is still missing rather than for what merely resembles the query.
To realize metamemory monitoring, ReMind carries a sufficiency predictor trained jointly with the ranking objective. At each step, the predictor estimates whether the memories collected so far are enough to answer the query, and terminates the loop when sufficiency is reached.
In this way, the number of selected memories is decided by the query itself instead of a fixed top-$K$: the loop runs further when the query needs more memories, and stops early when a few are enough. A simple query thus gets a short context with little noise and few tokens, and a complex query gets all the evidence it needs.
Our contributions can be summarized as follows:
\begin{itemize}[leftmargin=*]
    \item We identify two limitations shared by existing memory rankers: they fix the ranking in one pass and cut it at a fixed size, so the memories they select never inform the rest of the selection. 
    \item We propose ReMind, which reformulates memory ranking as a recurrent and self-terminating loop: it repeatedly promotes the top-ranked memory into the query context to rerank the remaining candidates, and terminates the loop through a sufficiency predictor.
    \item We conduct experiments on multiple public memory benchmarks, where ReMind achieves the best results compared to state-of-the-art memory rankers.
\end{itemize}

\section{Methodology}
\label{sec:methodology}

\begin{figure}[t]
    \centering
    \includegraphics[width=\linewidth]{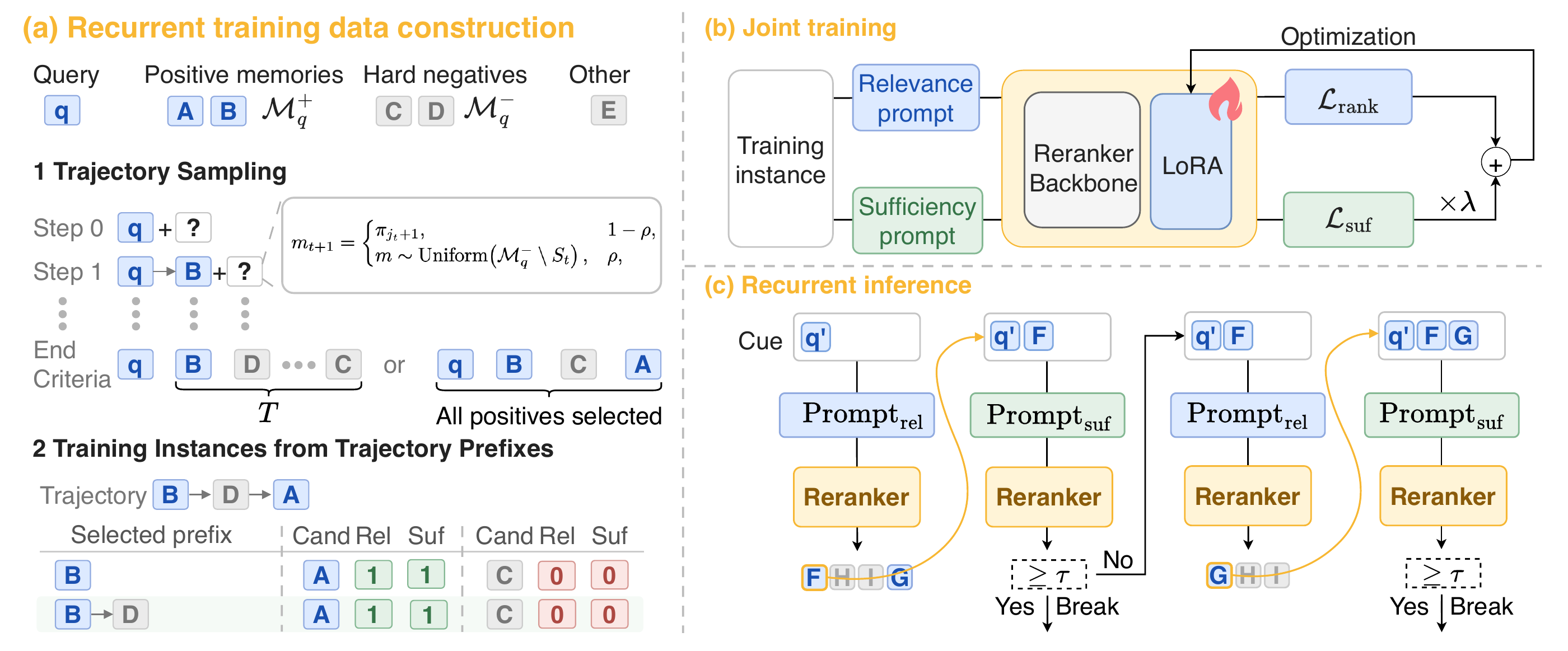}
    \caption{Overview of ReMind. Training examples are built from selection states that contain both positive and negative memories in the context. During inference, the model scores the remaining candidates conditioned on the selected memories, adds the top-ranked memory, and checks whether the updated context is sufficient.}
    \label{fig:remind-overview}
\end{figure}

As shown in Figure~\ref{fig:remind-overview}, ReMind combines a relevance predictor that conditions on selected memories with a sufficiency predictor that determines when to stop selection. We first describe how training examples are constructed from intermediate selection states in Section~\ref{sec:trajectory-data}. Section~\ref{sec:remind-model} presents the model inputs and the two prediction tasks, and Section~\ref{sec:remind-optimization} gives their joint training objective. Finally, Section~\ref{sec:remind-inference} describes how the two tasks work together during recurrent inference.

\subsection{Recurrent Training Data Construction}
\label{sec:trajectory-data}

\paragraph{Candidate preparation.}

A single-pass ranker is trained on instances $(q,m,y_m)$, where $q$ is the query, $m$ is a candidate memory, and $y_m$ is its binary relevance label. ReMind also conditions on the memories selected so far, denoted by $S_t$, so its training instances take the form $(q,S_t,m,y_m)$. We therefore construct training data that cover different intermediate selection states.
For each query $q$, the training data provide the set of positive memories $\mathcal{M}_q^+=\{m_1^+,\ldots,m_n^+\}$ that are needed to answer it.
Negative memories are not annotated, so we construct them from the retrieval stage that precedes ranking.
Given $q$, an upstream retriever returns the top-$K_r$ memories from the memory collection of the user as the candidate pool $\mathcal{M}_q$.
We take the hard negatives as the highest-ranked candidates that are not positive,
\begin{equation}
    \mathcal{M}_q^-=\operatorname{top}_{K_n}\!\left(\mathcal{M}_q\setminus\mathcal{M}_q^+\right),
    \label{eq:hard-negatives}
\end{equation}
where $\operatorname{top}_{K_n}(\cdot)$ keeps the $K_n$ memories with the highest retrieval scores.
These memories are ranked high by the retriever although they do not answer the query, so they are the ones the ranker must learn to reject.

\paragraph{Trajectory sampling.}
We describe a selection state by the ordered sequence of selected memories $S_t=(m_1,\ldots,m_t)$, where $S_0=()$ is the empty sequence.
The training data should cover states whose context mixes positive and negative memories, so we sample selection trajectories as follows.
For a query $q$, we first sample an ordering $\pi=(\pi_1,\ldots,\pi_n)$ of the positive memories in $\mathcal{M}_q^+$, where $\pi_i$ is the memory at position $i$ in the ordering.
Starting from $S_0$, the trajectory grows by one memory at each step according to
\begin{equation}
    m_{t+1}=
    \begin{cases}
        \pi_{j_t+1}, & \text{with probability } 1-\rho,\\
        m\sim\mathrm{Uniform}\!\left(\mathcal{M}_q^-\setminus \operatorname{set}(S_t)\right), & \text{with probability } \rho,
    \end{cases}
    \label{eq:trajectory-transition}
\end{equation}
where $j_t=|\operatorname{set}(S_t)\cap\mathcal{M}_q^+|$ is the number of positives already selected, so $\pi_{j_t+1}$ is the next unselected positive in the ordering.
The state is updated as $S_{t+1}=S_t\oplus m_{t+1}$, with $\oplus$ appending a memory to the sequence.
The negative ratio $\rho$ controls how often negative memories enter the context.
The trajectory ends once every positive in $\pi$ has been selected or a step limit $T$ is reached.
We repeat this procedure with up to $P$ distinct orderings per query.
Negative memories are placed in the context on purpose: at inference time the ranker may select a wrong memory, and it should still find the remaining positives when the context contains distractors.

\paragraph{Training instances from trajectory prefixes.}
Each prefix $S_t$ of a sampled trajectory is a selection state, that is, a context of selected memories that the ranker faces at inference time.
We build training instances at every state, each pairing $S_t$ with one candidate memory $m$ and a relevance label $y_m$.
The candidate is either a positive memory that is not yet in $S_t$, labeled $y_m=1$, or one of the hard negatives that are not yet in $S_t$, labeled $y_m=0$.
Negative memories thus serve two roles, as distractors inside the context and as candidates to be rejected, but never as selection targets, because supervising the ranker to select a negative memory would inject noise into the training signal.

\paragraph{Sufficiency labels.}
For the sufficiency predictor, each instance also needs a label that tells whether the collected memories are already enough to answer the query.
We derive this label from the annotated positives: the context after the candidate is added, $S_t\oplus m$, is sufficient if it contains every memory in $\mathcal{M}_q^+$\footnote{For datasets with interchangeable memories for the same evidence, sufficiency requires at least one memory from each evidence group.},
$z_{t,m}=\mathbf{1}\!\left[\mathcal{M}_q^+\subseteq \operatorname{set}(S_t\oplus m)\right]$.
The label depends on both the state and the candidate: it equals $1$ only when the candidate supplies the last missing positive, and $0$ otherwise.
With this label, a training instance becomes the tuple $(q,S_t,m,y_m,z_{t,m})$.

\subsection{ReMind Model}
\label{sec:remind-model}

\paragraph{Inputs across recurrent steps.}
A single-pass ranker scores a memory from the pair $(q,m)$.
ReMind additionally conditions on the ordered sequence of memories already selected, giving the relevance input
\begin{equation}
    x^{\mathrm{rel}}_{t,m}=\operatorname{Prompt}_{\mathrm{rel}}(q,S_t,m).
    \label{eq:recurrent-input}
\end{equation}
where $x^{\mathrm{rel}}_{t,m}$ is the input token sequence fed to the model.
At the first step, $S_0$ is empty and the input reduces to the query and the candidate.
After the top-ranked memory $m_{t+1}$ is selected, the context is updated to $S_{t+1}=S_t\oplus m_{t+1}$ and the remaining candidates are scored again under the expanded context.
The same model is applied at every step, and the growing memory sequence carries information from earlier selections into later ranking steps.
This is the recurrent component of ReMind: each selection changes the input of the next ranking step, so memories selected early serve as extra cues for memories that the raw query alone cannot match.
The relevance prompt uses the following fields, with its instruction shortened here for readability: 
  \begingroup
  \small
  \setlength{\leftmargini}{1em}
  \begin{quote}
  \raggedright
  \texttt{<Instruct>:} Does the candidate add evidence needed to answer
  the question?\\
  \mbox{\texttt{<Question>:} $q$};
  \mbox{\texttt{<Already Retrieved Memories>:} $S_t$};
  \mbox{\texttt{<Candidate Memory>:} $m$}
  \end{quote}
  \endgroup

\paragraph{Relevance and sufficiency prediction.}
ReMind does not depend on a particular reranker architecture.
In this paper, we build it on a language-model-based reranker that answers a relevance question with the tokens \texttt{yes} or \texttt{no}.
Given an input sequence $x$, the backbone produces a hidden representation at the final position, and the output projection maps it to the logits $\ell_\theta^{\mathrm{yes}}(x)$ and $\ell_\theta^{\mathrm{no}}(x)$ of the two answer tokens.
We normalize the two logits to obtain a probability
\begin{equation}
    p_\theta(x)=\frac{\exp\!\left(\ell_\theta^{\mathrm{yes}}(x)\right)}
    {\exp\!\left(\ell_\theta^{\mathrm{yes}}(x)\right)+\exp\!\left(\ell_\theta^{\mathrm{no}}(x)\right)},
    \qquad r_\theta(q,S_t,m)=p_\theta\!\left(x^{\mathrm{rel}}_{t,m}\right),
    \label{eq:binary-score}
\end{equation}
where $r_\theta$ is the relevance score used to rank the candidates at step $t$.
The sufficiency predictor reuses the same backbone and output projection through a second prompt, which asks whether the selected memories contain enough information to fully answer the query. The full prompts are given in \textbf{Appendix~\ref{app:prompts}}.
The sufficiency score is
\begin{equation}
    v_\theta(q,S)=p_\theta\!\left(\operatorname{Prompt}_{\mathrm{suf}}(q,S)\right),
    \qquad S=S_t\oplus m,
    \label{eq:sufficiency-score}
\end{equation}
where $S$ is the context after the candidate is added. 
Checking the updated context allows the predictor to detect whether the newly added memory completes the required evidence, so selection can stop before another ranking step.
The two tasks share all parameters and differ only in their prompts, so sufficiency prediction adds no parameters to the ranker.

\subsection{Optimization}
\label{sec:remind-optimization}

Following previous work~\citep{zhang2025qwen3}, which treats reranking as a binary classification task, we train relevance and sufficiency prediction jointly as two binary classification tasks on the constructed instances $(q,S_t,m,y_m,z_{t,m})$.
The ranking objective is as follows:
\begin{equation}
    \mathcal{L}_{\mathrm{rank}}
    =-\mathbb{E}_{\mathcal{D}}\!\left[y_m\log r_\theta(q,S_t,m)+(1-y_m)\log\!\left(1-r_\theta(q,S_t,m)\right)\right],
    \label{eq:ranking-loss}
\end{equation}
where $\mathcal{D}$ denotes the set of constructed training instances.
The sufficiency objective supervises the state after the candidate is added:
\begin{equation}
    \mathcal{L}_{\mathrm{suf}}
    =-\mathbb{E}_{\mathcal{D}}\!\left[z_{t,m}\log v_\theta(q,S_t\oplus m)+(1-z_{t,m})\log\!\left(1-v_\theta(q,S_t\oplus m)\right)\right].
    \label{eq:sufficiency-loss}
\end{equation}
The sufficiency target is binary: it indicates whether the required evidence is complete.
The two losses are combined as
\begin{equation}
    \mathcal{L}=\mathcal{L}_{\mathrm{rank}}+\lambda\mathcal{L}_{\mathrm{suf}},
    \label{eq:joint-loss}
\end{equation}
where $\lambda$ controls the weight of sufficiency supervision.
We train LoRA adapters~\citep{hu2021lora} on the shared backbone and keep the pretrained weights frozen.

\subsection{Recurrent Inference}
\label{sec:remind-inference}

\begin{wrapfigure}{r}{0.5\textwidth}
\vspace{-1.0\baselineskip}
\begin{minipage}{0.5\textwidth}
\setlength{\intextsep}{0pt}
\begin{algorithm}[H]
    \caption{Recurrent inference of ReMind}
    \label{alg:remind}
    \begin{algorithmic}[1]
        \Require Query $q$, candidates $\mathcal{M}_q$, threshold $\tau$
        \Ensure Selected memory sequence $S$
        \State $S\gets()$; $\mathcal{U}\gets\mathcal{M}_q$
        \While{$\mathcal{U}\neq\emptyset$ and budget not reached}
            \State $m^*\gets\operatorname*{arg\,max}_{m\in\mathcal{U}}r_\theta(q,S,m)$
            \State $S\gets S\oplus m^*$; $\mathcal{U}\gets\mathcal{U}\setminus\{m^*\}$
            \If{$v_\theta(q,S)\geq\tau$}
                \State \textbf{break}
            \EndIf
        \EndWhile
        \State \Return $S$
    \end{algorithmic}
\end{algorithm}
\end{minipage}
\end{wrapfigure}
At inference time, ReMind starts from the candidate pool $\mathcal{M}_q$ and the empty sequence $S_0$, and runs the loop summarized in Algorithm~\ref{alg:remind}.
At step $t$, it scores every remaining candidate conditioned on the current sequence and selects the highest-scoring memory,
\begin{equation}
    m_{t+1}=\operatorname*{arg\,max}_{m\in\mathcal{M}_q\setminus\operatorname{set}(S_t)}r_\theta(q,S_t,m),
    \label{eq:recurrent-selection}
\end{equation}
which is appended to the context, $S_{t+1}=S_t\oplus m_{t+1}$.
It then computes the sufficiency score $v_\theta(q,S_{t+1})$ on the updated context and stops if the score reaches a threshold $\tau$; otherwise, $S_{t+1}$ becomes the context of the next ranking step.
This check matches how the sufficiency labels are defined in training.
Selection also stops when the number of selected memories reaches a preset maximum or the candidate pool is exhausted.

\section{Experiments}
\label{sec:experiments}

\begin{table}[!t]
\caption{Main results. Bold marks the best result and underline marks the second best.}
\label{tab:main-dense}
\small
\centering
\setlength{\tabcolsep}{3pt}
\renewcommand{\arraystretch}{0.95}
\begin{tabular*}{\textwidth}{@{\hspace{2pt}\extracolsep{\fill}}l*{9}{c}@{\hspace{2pt}}}
\toprule
\textbf{Method} & \textbf{R@1} & \textbf{R@3} & \textbf{R@5} & \textbf{N@1} & \textbf{N@3} & \textbf{N@5} & \textbf{MAP} & \textbf{MRR} & \textbf{F1} \\
\midrule
\multicolumn{10}{@{}l@{}}{\makebox[0pt][l]{\colorbox[gray]{0.95}{\makebox[\dimexpr\textwidth-2\fboxsep\relax]{\strut\textbf{\textit{LongMemEval-M}}}}}} \\
Base & 24.62 & 46.40 & 53.83 & 43.79 & 44.24 & 47.71 & 45.80 & 54.64 & 33.74 \\
\midrule
Qwen3-Reranker-0.6B~\citep{zhang2025qwen3} & 39.73 & 64.88 & 70.52 & 68.64 & 64.77 & 67.19 & 64.15 & 75.93 & 47.50 \\
Qwen3-Reranker-4B~\citep{zhang2025qwen3} & 45.08 & 73.14 & 79.89 & 75.74 & 74.19 & 76.50 & 72.99 & 81.65 & 55.06 \\
Qwen3-Reranker-8B~\citep{zhang2025qwen3} & 44.85 & 73.42 & 80.83 & 76.33 & 74.45 & 77.35 & 73.72 & 82.18 & 54.74 \\
KaLM-Nano~\citep{zhao2026kalm} & 39.69 & 66.77 & 74.96 & 70.41 & 67.56 & 70.84 & 67.34 & 76.91 & 50.04 \\
KaLM-Small~\citep{zhao2026kalm} & 48.32 & 75.90 & 79.34 & 84.02 & 78.49 & 79.24 & 76.73 & 86.39 & 57.29 \\
KaLM-Large~\citep{zhao2026kalm} & 48.18 & 75.70 & 81.39 & 82.84 & 78.01 & 79.81 & 76.77 & 86.08 & 56.87 \\
RankGPT~\citep{sun2023chatgpt} & 43.72 & 70.70 & 73.75 & 76.92 & 72.23 & 72.67 & 70.42 & 80.69 & 53.22 \\
\midrule
MemReranker-0.6B~\citep{li2026memreranker} & 46.08 & 76.12 & 80.93 & 79.88 & 77.66 & 79.11 & 76.13 & 84.49 & 57.57 \\
MemReranker-4B~\citep{li2026memreranker} & \underline{49.21} & \underline{79.31} & \underline{82.41} & \underline{85.80} & \underline{81.81} & \underline{82.14} & \underline{79.57} & \underline{87.70} & \underline{60.18} \\
QRRanker~\citep{li2026query} & 35.65 & 63.63 & 70.78 & 64.50 & 63.29 & 66.02 & 62.74 & 73.07 & 47.73 \\
ConvMemory~\citep{pan2026convmemory,pan2026convmemoryv2} & 34.95 & 59.44 & 69.23 & 60.36 & 58.77 & 63.30 & 59.53 & 69.54 & 43.89 \\
\midrule
\textbf{ReMind-SFT} & 49.03 & 77.35 & 80.79 & 85.21 & 79.99 & 80.63 & 77.95 & 87.31 & 58.37 \\
\textbf{ReMind} & \textbf{50.07} & \textbf{80.53} & \textbf{82.82} & \textbf{86.39} & \textbf{83.17} & \textbf{83.03} & \textbf{80.74} & \textbf{88.41} & \textbf{60.87} \\
\midrule
\multicolumn{10}{@{}l@{}}{\makebox[0pt][l]{\colorbox[gray]{0.95}{\makebox[\dimexpr\textwidth-2\fboxsep\relax]{\strut\textbf{\textit{PersonaBench}}}}}} \\
Base & 17.41 & 33.07 & 42.47 & 30.86 & 33.28 & 35.90 & 36.52 & 43.54 & 22.67 \\
\midrule
Qwen3-Reranker-0.6B~\citep{zhang2025qwen3} & 24.47 & 38.91 & 53.27 & 38.27 & 39.98 & 44.64 & 43.14 & 51.25 & 26.17 \\
Qwen3-Reranker-4B~\citep{zhang2025qwen3} & 35.70 & 60.78 & 70.37 & 54.32 & 60.41 & 62.64 & 59.39 & 65.88 & 40.37 \\
Qwen3-Reranker-8B~\citep{zhang2025qwen3} & 39.20 & 60.04 & 69.96 & 56.79 & 61.20 & 63.01 & 61.03 & 66.94 & 40.37 \\
KaLM-Nano~\citep{zhao2026kalm} & 22.45 & 46.54 & 58.17 & 38.27 & 44.90 & 47.88 & 45.55 & 52.64 & 31.31 \\
KaLM-Small~\citep{zhao2026kalm} & 31.38 & 58.37 & 69.07 & 50.62 & 56.47 & 58.70 & 55.52 & 63.17 & 38.35 \\
KaLM-Large~\citep{zhao2026kalm} & 37.67 & 54.63 & 70.35 & 58.02 & 58.57 & 62.30 & 59.65 & 66.46 & 38.02 \\
RankGPT~\citep{sun2023chatgpt} & 33.79 & 52.04 & 61.03 & 53.09 & 54.25 & 54.81 & 53.95 & 62.02 & 35.99 \\
\midrule
MemReranker-0.6B~\citep{li2026memreranker} & 31.19 & 49.96 & 64.01 & 46.91 & 51.03 & 55.13 & 53.24 & 58.47 & 34.17 \\
MemReranker-4B~\citep{li2026memreranker} & 33.48 & 51.30 & 61.73 & 53.09 & 54.35 & 56.09 & 55.47 & 62.22 & 35.78 \\
QRRanker~\citep{li2026query} & 26.07 & 50.80 & 65.08 & 44.44 & 49.42 & 53.64 & 50.44 & 58.96 & 33.62 \\
ConvMemory~\citep{pan2026convmemory,pan2026convmemoryv2} & 17.59 & 32.41 & 40.88 & 27.16 & 32.60 & 34.52 & 35.26 & 40.24 & 22.91 \\
\midrule
\textbf{ReMind-SFT} & \underline{41.77} & \underline{65.00} & \underline{75.84} & \underline{59.26} & \underline{66.08} & \underline{67.84} & \underline{65.50} & \underline{70.83} & \underline{43.58} \\
\textbf{ReMind} & \textbf{49.81} & \textbf{71.30} & \textbf{78.74} & \textbf{72.84} & \textbf{74.77} & \textbf{74.28} & \textbf{72.61} & \textbf{79.25} & \textbf{48.09} \\
\midrule
\multicolumn{10}{@{}l@{}}{\makebox[0pt][l]{\colorbox[gray]{0.95}{\makebox[\dimexpr\textwidth-2\fboxsep\relax]{\strut\textbf{\textit{ConvoMem}}}}}} \\
Base & 21.00 & 48.78 & 61.82 & 68.00 & 62.06 & 62.54 & 61.46 & 78.81 & 50.06 \\
\midrule
Qwen3-Reranker-0.6B~\citep{zhang2025qwen3} & 23.09 & 57.46 & 69.72 & 75.50 & 72.15 & 71.31 & 69.94 & 84.31 & 58.91 \\
Qwen3-Reranker-4B~\citep{zhang2025qwen3} & 25.72 & 66.43 & 83.01 & 85.50 & 83.80 & 83.90 & 82.68 & 91.02 & 69.03 \\
Qwen3-Reranker-8B~\citep{zhang2025qwen3} & 25.31 & 63.73 & 79.92 & 84.00 & 80.87 & 81.23 & 80.27 & 90.07 & 66.20 \\
KaLM-Nano~\citep{zhao2026kalm} & 27.18 & 63.63 & 79.37 & 88.00 & 81.33 & 81.40 & 80.29 & 91.80 & 65.75 \\
KaLM-Small~\citep{zhao2026kalm} & 27.82 & 67.21 & 82.41 & 90.50 & 85.01 & 84.68 & 83.42 & 93.63 & 69.37 \\
KaLM-Large~\citep{zhao2026kalm} & 27.72 & 67.53 & 83.21 & 90.00 & 85.52 & 85.34 & 83.93 & 93.88 & 69.74 \\
RankGPT~\citep{sun2023chatgpt} & 24.38 & 61.92 & 74.52 & 79.50 & 77.89 & 76.34 & 76.16 & 86.97 & 63.98 \\
\midrule
MemReranker-0.6B~\citep{li2026memreranker} & 25.66 & 63.19 & 78.95 & 84.50 & 80.06 & 80.35 & 79.36 & 90.33 & 65.29 \\
MemReranker-4B~\citep{li2026memreranker} & 25.23 & 63.76 & 80.46 & 81.50 & 79.78 & 80.94 & 80.28 & 88.73 & 65.80 \\
QRRanker~\citep{li2026query} & 21.54 & 55.73 & 72.86 & 70.50 & 69.41 & 72.03 & 70.95 & 81.06 & 57.16 \\
ConvMemory~\citep{pan2026convmemory,pan2026convmemoryv2} & 15.19 & 38.99 & 51.48 & 45.50 & 46.33 & 49.08 & 50.21 & 62.41 & 38.92 \\
\midrule
\textbf{ReMind-SFT} & \underline{28.97} & \underline{75.00} & \underline{92.43} & \underline{97.00} & \underline{95.74} & \underline{95.20} & \underline{93.62} & \underline{98.33} & \underline{78.68} \\
\textbf{ReMind} & \textbf{29.97} & \textbf{76.75} & \textbf{93.77} & \textbf{99.00} & \textbf{97.51} & \textbf{96.75} & \textbf{95.10} & \textbf{99.50} & \textbf{80.08} \\
\bottomrule
\end{tabular*}
\end{table}

\subsection{Experimental Setup}
\label{sec:experimental-setup}

\paragraph{Datasets and Evaluation Metrics.}
We evaluate ReMind on four memory benchmarks: LongMemEval-S and LongMemEval-M \citep{wu2024longmemeval}, PersonaBench \citep{tan2025personabench}, and ConvoMem \citep{pakhomov2025convomem}.
We choose these benchmarks because each of them annotates the memories in the user history that are needed to answer every query.
Following prior work on memory retrieval and reranking~\citep{wu2024longmemeval,li2026memreranker}, we report Recall@$K$ (R@$K$) and NDCG@$K$ (N@$K$) with $K\in\{1,3,5\}$, MAP, MRR, and F1.
Please refer to \textbf{Appendix~\ref{app:dataset}} for detailed dataset introduction.

\paragraph{Baselines.}
We compare ReMind with two groups of rerankers, and also report the original retrieval order (Base).
The first group consists of general-purpose rerankers that are not specialized for memory ranking: Qwen3-Reranker~\citep{zhang2025qwen3} with 0.6B, 4B, and 8B parameters, KaLM-Reranker-V1~\citep{zhao2026kalm} in its Nano, Small, and Large versions, and RankGPT~\citep{sun2023chatgpt}, which prompts a large language model to rerank the candidate list without training.
The second group consists of rerankers designed specifically for memory ranking: MemReranker~\citep{li2026memreranker} with 0.6B and 4B parameters, QRRanker~\citep{li2026query}, and ConvMemory-LoCoMo-MPNet~\citep{pan2026convmemory,pan2026convmemoryv2}. We also include ReMind-SFT, a single-pass reranker trained from the same backbone on the same queries with the standard reranker recipe. For more details on the baselines, please refer to \textbf{Appendix~\ref{app:baselines}}.

\paragraph{Implementation Details.}
To ensure a fair comparison, all rerankers share exactly the same frozen top-50 candidate set returned by a retriever, so comparisons focus on ranking and memory selection after retrieval.
ReMind is initialized from Qwen3-Reranker-4B~\citep{zhang2025qwen3}, and other scales are studied in \textbf{Appendix~\ref{app:backbone-scales}}.
For trajectory construction, we use $K_n=4$ hard negatives, $P=5$ positive orderings per query, and a negative ratio of $\rho=0.2$, and its design choices are studied in \textbf{Appendix~\ref{app:trajectory-construction}}.
For LoRA fine-tuning~\citep{hu2021lora}, we apply adapters to all linear projection layers in the attention and feed-forward blocks, with a rank of 16, a scaling factor of 32, and a dropout of 0.05.
For joint optimization, we set the joint-loss weight $\lambda=0.6$.
At inference, the sufficiency threshold is $\tau=0.5$.
Additional settings are provided in \textbf{Appendix~\ref{app:key-parameters}} for reference.

\subsection{Main Results}
\label{sec:main-results}

We run experiments with three retrievers and four datasets.
The retrievers cover a sparse retriever, BM25~\citep{robertson2009probabilistic}; a dense retriever\footnote{\url{https://huggingface.co/sentence-transformers/multi-qa-MiniLM-L6-cos-v1}}; and a state-of-the-art memory retriever, RF-Mem~\citep{zhang2026evoking}.
Because of space limits, the main text reports the results with the dense retriever, which is the most common choice in memory systems, on LongMemEval-M, PersonaBench, and ConvoMem in Table~\ref{tab:main-dense}. The complete results are given in \textbf{Appendix~\ref{app:complete-main-results}}.
For a fair comparison, ReMind is evaluated with the same number of memories as the other rankers.
When ReMind stops before the cutoff because the sufficiency predictor activates, the remaining positions are filled with the unselected candidates ordered by their relevance scores from the last step; when it reaches the selection limit without stopping, the ranking is truncated at the cutoff. Three observations follow from Table~\ref{tab:main-dense}.
First, ReMind achieves the strongest overall performance across benchmarks and retrievers, which supports the effectiveness of the recurrent selection mechanism.
Second, the gains are larger on PersonaBench and ConvoMem than on LongMemEval-M.
One possible reason is that queries in PersonaBench and ConvoMem often need several memories that are individually only weakly related to the query, so the memories already selected provide cues for finding the remaining ones.
Third, the improvement does not come only from training on the target data.
On LongMemEval-M, ReMind-SFT stays below MemReranker-4B on all nine metrics, while ReMind exceeds it on all nine.
On PersonaBench, the recurrent selection adds a large improvement over ReMind-SFT.
On ConvoMem, where ReMind-SFT already reaches 95.74 NDCG@3, the recurrent selection still removes about 40\% of the remaining error.

\subsection{Effectiveness of the Sufficiency Predictor}
\label{sec:sufficiency}

\begin{wrapfigure}{r}{0.5\textwidth}
\vspace{-1.0\baselineskip}
\centering
\includegraphics[width=0.48\textwidth]{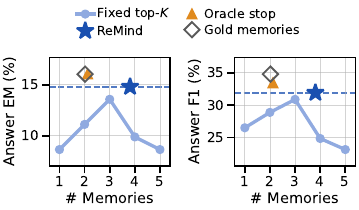}
\caption{Impact of sufficiency predictor on downstream question answering. Fixed top-$K$ runs the ReMind loop for $K$ steps. Oracle stop ranks with ReMind and stops once all gold memories have been selected, and Gold memories places the gold set in the context directly.}
\label{fig:sufficiency}
\vspace{-0.5\baselineskip}
\end{wrapfigure}
We examine whether the number of memories chosen by the sufficiency predictor is better than any fixed cutoff.
We pass the memories selected by ReMind to a reader (Llama-3-8B-Instruct~\citep{grattafiori2024llama}) that answers the query, and compare sufficiency stopping with fixed cutoffs of one to five memories (running the ReMind loop for $K$ steps without the sufficiency predictor) on PersonaBench with the dense retriever.
Figure~\ref{fig:sufficiency} reports exact match and answer F1, and the complete results are given in \textbf{Appendix~\ref{app:sufficiency}}.
We can draw the following conclusions.
First, answer quality does not grow with the number of memories: it peaks at three and then falls.
One possible reason is that the memories added after the third one are rarely needed for the answer, so they act as distractors for the reader.
Second, no fixed cutoff reaches the accuracy of sufficiency stopping at any budget, which proves the effectiveness of sufficiency predictor.
Third, the headroom that remains lies in stopping rather than in ranking, as stopping the same ReMind ranking as soon as all gold memories have been selected reaches the same exact match as placing the gold memory set in the context directly.
Running the loop also adds ranking time, which remains small compared with the time the reader needs to generate an answer, as analyzed in \textbf{Appendix~\ref{app:inference-cost}}.

\subsection{Ablation Study}
\label{sec:ablation}

\begin{table}[t]
\caption{Ablation study on PersonaBench with dense retrieval. RT is recurrent training, SL is the sufficiency loss, RI is recurrent inference, and SS is sufficiency stopping. Bold marks the best value in each metric column and underline marks the second best.}
\label{tab:ablation}
\small
\centering
\setlength{\tabcolsep}{3pt}
\renewcommand{\arraystretch}{1.0}
\begin{tabular*}{\textwidth}{@{\hspace{2pt}\extracolsep{\fill}}c*{4}{c}*{10}{c}@{\hspace{2pt}}}
\toprule
\textbf{ID} & \textbf{RT} & \textbf{SL} & \textbf{RI} & \textbf{SS} & \textbf{R@1} & \textbf{R@3} & \textbf{R@5} & \textbf{N@1} & \textbf{N@3} & \textbf{N@5} & \textbf{MAP} & \textbf{MRR} & \textbf{F1} \\
\midrule
1 & \no & \no & \no & \no & 41.77 & 65.00 & 75.84 & 59.26 & 66.08 & 67.84 & 65.50 & 70.83 & 43.58 \\
2 & \no & \no & \yes & \no & 41.77 & 59.40 & 72.14 & 59.26 & 62.56 & 65.33 & 64.04 & 69.17 & 40.39 \\
3 & \yes & \no & \no & \no & \underline{42.39} & 65.62 & 77.70 & \underline{60.49} & 66.84 & 69.13 & 66.32 & 72.24 & 44.07 \\
4 & \yes & \no & \yes & \no & \underline{42.39} & 67.88 & 78.11 & \underline{60.49} & 68.94 & 70.42 & 67.56 & 73.95 & 45.60 \\
5 & \yes & \yes & \no & \no & \textbf{49.81} & 66.36 & 76.28 & \textbf{72.84} & 71.67 & 72.17 & 71.01 & 77.86 & 45.62 \\
\midrule
6 & \yes & \yes & \yes & \no & \textbf{49.81} & \textbf{71.60} & \textbf{79.92} & \textbf{72.84} & \textbf{75.06} & \textbf{75.01} & \textbf{73.06} & \textbf{79.62} & \textbf{48.44} \\
7 & \yes & \yes & \yes & \yes & \textbf{49.81} & \underline{71.30} & \underline{78.74} & \textbf{72.84} & \underline{74.77} & \underline{74.28} & \underline{72.61} & \underline{79.25} & \underline{48.09} \\
\bottomrule
\end{tabular*}
\end{table}

To verify that each component of ReMind contributes to its performance, we conduct an ablation analysis over four design choices.
Recurrent training (RT) builds the training instances from selection trajectories, so the ranker is supervised at states where some memories have already been selected; without it the backbone is trained on one positive memory and several hard negatives per query.
The sufficiency loss (SL) adds the sufficiency prediction task to the training objective.
Recurrent inference (RI) selects memories one at a time and rescores the remaining candidates with the selected ones in the context, instead of ranking all candidates in a single pass.
Sufficiency stopping (SS) ends the selection as soon as the sufficiency predictor activates, instead of always reaching the selection limit.
Table~\ref{tab:ablation} reports the combinations we study on PersonaBench with the dense retriever.
We can draw the following conclusions.
First, the recurrent design has to be present in both training and inference: recurrent training alone raises NDCG@3 by 0.76 (rows 1 and 3) and recurrent inference alone lowers it by 3.52 (rows 1 and 2), while the two together raise it by 2.86 (rows 1 and 4).
Second, the sufficiency loss and recurrent inference reinforce each other instead of acting independently: the sufficiency loss adds 4.83 NDCG@3 without recurrent inference (rows 3 and 5) and 6.12 with it (rows 4 and 6), while recurrent inference adds 2.10 without the sufficiency loss (rows 3 and 4) and 3.39 with it (rows 5 and 6). 
Third, sufficiency stopping is not meant to improve the ranking metrics. For a fair ranking comparison, every run is evaluated on the same number of memories, so when the loop stops early, the remaining positions are filled with the unselected candidates ordered by their scores from the last step instead of by further recurrent steps. The benefit of sufficiency stopping lies in the downstream task as shown in Section~\ref{sec:sufficiency}.

\subsection{Hyperparameter Analysis}
\label{sec:hyperparameter}

\begin{figure}[h]
\centering
\includegraphics[width=\textwidth]{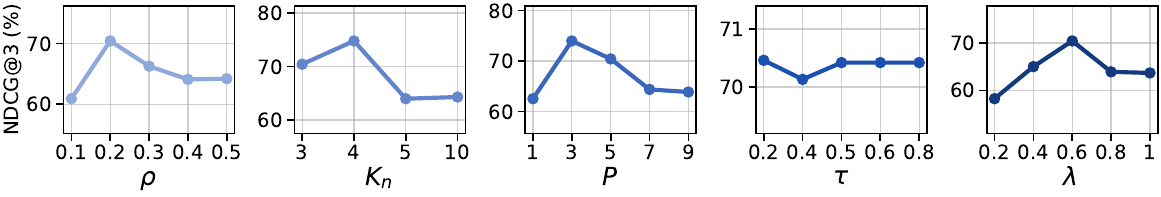}
\vspace{-8mm}
\caption{Hyperparameter analysis on PersonaBench with dense retrieval.}
\label{fig:hyperparameter}
\end{figure}

We vary one hyperparameter at a time while keeping the others fixed at the starting values of this analysis, namely negative ratio $\rho=0.2$, hard-negative count $K_n=3$, positive orderings per query $P=5$, sufficiency threshold $\tau=0.5$, and joint-loss weight $\lambda=0.6$, and the other experiments use the settings given in Section~\ref{sec:experimental-setup}.
Figure~\ref{fig:hyperparameter} reports NDCG@3 on PersonaBench, with similar trends across metrics.
Complete results appear in \textbf{Appendix~\ref{app:complete-hyperparameter-results}}.

\textbf{Data construction.}
(1) Performance first improves and then declines as $\rho$ increases.
Moderate negative exposure helps the ranker recover from incorrect selections, but excessive exposure may create training contexts unlike the ones seen at inference.
(2) Increasing $K_n$ shows a similar trend: a larger pool adds diverse hard negatives, but lower-ranked candidates tend to be easier to reject and may weaken supervision.
(3) Performance peaks at $P=3$.
Queries with one or two positives allow at most one or two distinct orderings, respectively.
Larger $P$ therefore gives more weight to queries with several positives and their later selection states, potentially shifting training away from queries with few positives, which make up a large part of datasets.

\textbf{Training and inference.}
(1) Performance remains stable for $\tau$ between 0.2 and 0.8, suggesting that ranking quality is not sensitive to the precise sufficiency threshold within this range.
(2) Performance first improves and then declines as $\lambda$ increases.
Moderate sufficiency supervision supports evidence-completeness prediction, while excessive weight may compromise relevance learning.

\subsection{Performance by Query Complexity}
\label{sec:query-complexity}

\begin{wraptable}{r}{0.45\textwidth}
\vspace{-1.0\baselineskip}
\caption{NDCG@3 by the number of gold memories a query needs}
\label{tab:query-complexity}
\small
\centering
\setlength{\tabcolsep}{3pt}
\renewcommand{\arraystretch}{1.0}
\begin{tabular*}{\linewidth}{@{\extracolsep{\fill}}lccc@{}}
\toprule
\textbf{Method} & \textbf{2} & \textbf{3} & \textbf{4+} \\
\midrule
MemReranker-4B & 56.45 & \underline{66.67} & 63.75 \\
ReMind-SFT & \underline{60.22} & 62.76 & \underline{69.07} \\
\textbf{ReMind} & \textbf{63.99} & \textbf{69.55} & \textbf{81.74} \\
\bottomrule
\end{tabular*}
\vspace{-0.5\baselineskip}
\end{wraptable}
We examine whether the advantage of ReMind depends on how many memories a query needs.
We group the PersonaBench queries by the number of gold memories into those that need two, three, and four or more, and we report NDCG@3 with the dense retriever for ReMind, ReMind-SFT, and MemReranker-4B, the strongest reranker designed for memory retrieval on this dataset.
Table~\ref{tab:query-complexity} reports the results, and Recall@3 is added in \textbf{Appendix~\ref{app:query-complexity}}.
We can draw the following conclusions.
First, ReMind is the best method in every group, so its advantage is not limited to one kind of query.
Second, the advantage is largest on the hardest group: on queries that need four or more memories, ReMind is 12.67 above ReMind-SFT and 17.99 above MemReranker-4B, while on queries that need two memories the two margins are 3.77 and 7.54.
Third, against ReMind-SFT, which shares the backbone, the margin grows with the number of required memories, from 3.77 to 6.79 and then to 12.67.
This supports the motivation of the recurrent design, because a query that needs more memories offers more selected memories to condition on.

\section{Related Work}

\textbf{Ranking in Information Retrieval.} Information retrieval commonly uses an efficient retriever to collect candidates and a more expensive ranker to refine their order~\citep{nogueira2019multistage,lin2021pretrained,guo2020deep,zhu2025large}.
Among pretrained neural rankers, monoBERT classifies query-candidate pairs with BERT~\citep{nogueira2019passage}, while monoT5 generates relevance labels with a sequence-to-sequence model~\citep{nogueira2020document}.
LLM-based methods either generate candidate rankings through prompting, as in RankGPT~\citep{sun2023chatgpt}, or learn to score query-candidate pairs, as in RankLLaMA~\citep{ma2023finetuning}, Qwen3-Reranker~\citep{zhang2025qwen3}, and KaLM-Reranker~\citep{zhao2026kalm}.

\textbf{Agent Memory.} Agent memory systems store and retrieve interaction history to support dialogue across sessions~\citep{zhong2024memorybank,chhikara2025mem0}.
For memory construction, MemoryBank combines conversations with summaries~\citep{zhong2024memorybank}, while SeCom compresses topical segments for retrieval~\citep{pan2025secom}.
A-Mem links related notes~\citep{xu2026mem}, Mem0 merges extracted facts with existing memories~\citep{chhikara2025mem0}, and MemInsight adds attributes to guide retrieval~\citep{salama2025meminsight}.
For retrieval, LD-Agent combines semantic, topical, and temporal signals~\citep{li2025hello}, while RF-Mem combines one-shot retrieval with stepwise recollection~\citep{zhang2026evoking}.

Closer to our work, MemReranker distills LLM judgments for memory ranking~\citep{li2026memreranker}, and QRRanker uses selected attention heads to score candidates jointly~\citep{li2026query}.
ConvMemory uses lightweight scoring with an optional cross-encoder refinement stage~\citep{pan2026convmemory,pan2026convmemoryv2}, while RMM updates its reranker through reinforcement learning from LLM citations~\citep{tan2025prospect}.
ReMind conditions each ranking step on previously selected memories and learns when to stop selection, complementing methods for memory construction and retrieval.

\section{Conclusion}
\label{sec:conclusion}

In this paper, we propose ReMind, reformulating memory ranking as a recurrent and self-terminating loop. ReMind uses selected memories as cues to rerank the remaining candidates and a shared-backbone sufficiency predictor to decide when to stop. 
Experiments on public memory benchmarks show that ReMind outperforms general-purpose and memory-specific rerankers.
One limitation is that the sufficiency predictor still stops later than it needs to, as in Section~\ref{sec:sufficiency}.
Closing this gap, for example by calibrating the sufficiency score or by adapting the stopping threshold to the query, is left for future work.

\clearpage

\subsection*{AI use statement}

We used generative AI tools in polishing the writing of this paper, and to assist with the implementation of the experimental code.
All text and code produced with such assistance were reviewed by the authors, and the code was tested before it was used to produce the reported results.
The authors take responsibility for the final content of this work, including the material produced with the aid of generative AI.

\subsection*{Reproducibility statement}

Implementation details are provided in Section~\ref{sec:experimental-setup}, with additional training and inference settings in Appendix~\ref{app:key-parameters}.

\bibliography{iclr2027_conference}
\bibliographystyle{iclr2027_conference}

\clearpage
\appendix
\begin{center}
    \LARGE \textbf{Technical Appendix}
    \vspace{1em}
\end{center}
\addcontentsline{toc}{part}{Appendix}

\begingroup
  \etocsetnexttocdepth{subsection}
  \etocsettocstyle{\section*{Table of Contents}}{}
  \localtableofcontents
\endgroup
\clearpage

\section{Prompt Templates}
\label{app:prompts}
ReMind uses two prompts that share the same chat format: a system message, a user message with labeled fields, and an assistant turn in which the model answers \texttt{yes} or \texttt{no}.
The relevance prompt scores one candidate memory given the query and the memories already selected, and the sufficiency prompt judges whether the selected memories are enough to answer the query.
The texts below are the templates used in our implementation.

\paragraph{Relevance prompt.}
\begin{quote}\small
\texttt{[System]} You are an iterative memory reranking assistant. Given a user question, the memories that have already been selected, and one candidate memory, decide whether the candidate provides additional information needed to answer the question. Answer only yes or no.\\[4pt]
\texttt{[User]}\\
\texttt{<Instruct>:} Select the candidate memory if it adds new relevant evidence for answering the user question beyond the already selected memories.\\
\texttt{<Question>:} \{query\}\\
\texttt{<Already Retrieved Memories>:} \{selected memories\}\\
\texttt{<Candidate Memory>:} \{candidate memory\}
\end{quote}

\paragraph{Sufficiency prompt.}
\begin{quote}\small
\texttt{[System]} You are a memory coverage judge. Decide whether the selected memories contain enough information to fully answer the user question. Answer only yes or no.\\[4pt]
\texttt{[User]}\\
\texttt{<Instruct>:} Answer yes if the selected memories are sufficient to fully answer the user question. Otherwise answer no.\\
\texttt{<Question>:} \{query\}\\
\texttt{<Selected Memories>:} \{selected memories\}
\end{quote}

\section{Introduction of Datasets}
\label{app:dataset}

This section introduces the datasets used in our experiments. The number of queries with annotated gold memories is limited in this setting, and prior work trains at a comparable scale~\citep{tan2025prospect}. We follow the same strategy and keep the backbone frozen, training only the LoRA adapters, because full fine-tuning needs more labeled data than this setting provides. The trajectory construction described in Section~\ref{sec:trajectory-data} also increases the supervision that a single query provides, because it builds one training instance for every pair of a selection state and a candidate, and it samples $P$ different orderings of the positive memories of each query. A query therefore turns into tens of training instances, and the different orderings act as data augmentation by presenting the same memories in different selection states.

\begin{itemize}[leftmargin=*,itemsep=3pt,parsep=0pt]
    \item \textbf{LongMemEval}~\citep{wu2024longmemeval}. LongMemEval evaluates the long-term memory of chat assistants with manually written questions that cover five abilities: information extraction, multi-session reasoning, temporal reasoning, knowledge updates, and abstention. Each question is answered from one or more task-oriented sessions between a user and an assistant, where the needed information is mentioned in passing rather than stated plainly, and most questions draw their evidence from several sessions. The evidence sessions are then placed among a large number of unrelated sessions to form one long chat history. We use the two standard settings released with the benchmark, LongMemEval-S with about 115k tokens per question and LongMemEval-M with 500 sessions per question. We use 60\% of the questions for training and validation, 40\% for testing, and the split is based on the question identifiers provided by the benchmark.

    \item \textbf{PersonaBench}~\citep{tan2025personabench}. PersonaBench measures how well a model understands a user by reading that user's private data. Its data come from synthetic user profiles that cover demographic, psychographic, and social attributes and that are linked to one another in social communities, and the profiles themselves are never shown to the model. Each profile generates three kinds of documents, namely conversations with connected people, interactions with an AI assistant, and purchase histories, in which the attributes surface indirectly and are mixed with sessions that carry no personal information. The questions ask about basic information, preferences, and social relations, and the social questions may take more than one step, such as identifying a relative before recalling what that person likes. We split the dataset into training, validation, and test sets by the personas in a 3:1:2 ratio.

    \item \textbf{ConvoMem}~\citep{pakhomov2025convomem}. ConvoMem is a large-scale benchmark for conversational memory with 75,336 question-answer pairs in six categories: user facts, assistant facts, abstention, preferences, changing facts, and implicit connections. The evidence for a question is spread over one to six messages, so a system has to collect several pieces of the conversation history instead of locating a single statement. The conversations that hold the evidence and the filler conversations come from the same generation pipeline, which prevents a system from spotting the evidence by its writing style. We sample 600 queries similar to PersonaBench and split the dataset into training, validation, and test sets by the personas in a 3:1:1 ratio.
\end{itemize}

Some of these benchmarks contain queries that the user history cannot answer, and such a query has no annotated gold memory.
We remove those queries, because a query with no gold memory cannot measure whether a ranker finds the right memories.
This also places them outside the scope of ReMind: the loop selects a memory before the first sufficiency check, so it always returns at least one memory and cannot report that the history holds no evidence.
Deciding that no memory in the pool can answer the query is a different problem from the one the sufficiency predictor solves, which is whether the memories selected so far are already enough, so we leave it to future work.

\section{Introduction of Baselines}
\label{app:baselines}

This section introduces the baselines used in our experiments.

\begin{itemize}[leftmargin=*,itemsep=3pt,parsep=0pt]
    \item \textbf{Base.} This baseline preserves the original order returned by the upstream retriever without applying a reranker. It uses the same candidate pool as the reranking methods.

    \item \textbf{Qwen3-Reranker}~\citep{zhang2025qwen3}. Qwen3-Reranker scores each query-document pair with a Qwen3 language model, using a task instruction to define relevance. It treats relevance as a binary decision and computes the score by normalizing the logits of the \texttt{yes} and \texttt{no} tokens. We evaluate the 0.6B, 4B, and 8B variants.

    \item \textbf{KaLM-Reranker-V1}~\citep{zhao2026kalm}. KaLM-Reranker-V1 uses an encoder to precompute passage representations and a decoder to score their relevance to the query through cross-attention. Matryoshka embedding pooling compresses the passage representations, allowing them to be stored and reused across queries. We evaluate the Nano, Small, and Large variants.

    \item \textbf{RankGPT}~\citep{sun2023chatgpt}. RankGPT prompts an LLM with a query and a list of candidate passages, asking it to output their identifiers in relevance order without task-specific training. Its sliding-window strategy handles long candidate lists by reranking overlapping groups of passages from the end of the list toward the beginning.

    \item \textbf{MemReranker}~\citep{li2026memreranker}. MemReranker adapts Qwen3-Reranker to agent memory retrieval using general retrieval data and memory-specific dialogue data. Its training first distills calibrated relevance labels from LLM teachers with a pointwise binary cross-entropy objective, then applies contrastive learning to distinguish hard candidates. We evaluate the 0.6B and 4B variants.

    \item \textbf{QRRanker}~\citep{li2026query}. QRRanker processes the candidate passages and query together, and derives continuous relevance scores from the attention weights between query tokens and passage tokens in selected attention heads. It trains these heads for listwise reranking, so each passage is scored with access to the other candidates in the input.

    \item \textbf{ConvMemory}~\citep{pan2026convmemory}. ConvMemory combines dense embeddings and lexical interaction features with a small convolutional/mixer network to score candidate memories. It is trained with cross-encoder teacher supervision on conversational memory retrieval data. We use the ConvMemory-LoCoMo-MPNet checkpoint.

    \item \textbf{ReMind-SFT}. ReMind-SFT starts from the same backbone as ReMind and is trained on the same queries with the standard reranker recipe, where each query is paired with one positive memory and several hard negatives. It ranks all candidates in a single pass, so it measures how much of the improvement comes from training on the target data rather than from the recurrent selection mechanism.
\end{itemize}

\section{Impact of Backbone Scales}
\label{app:backbone-scales}

\begin{figure}[t]
\begin{minipage}[t]{0.48\textwidth}
\centering
\includegraphics[width=\linewidth]{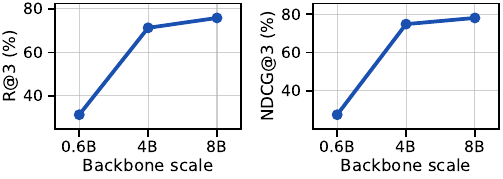}
\caption{Impact of backbone scales.}
\label{fig:backbone-scale}
\end{minipage}\hfill
\begin{minipage}[t]{0.48\textwidth}
\centering
\includegraphics[width=\linewidth]{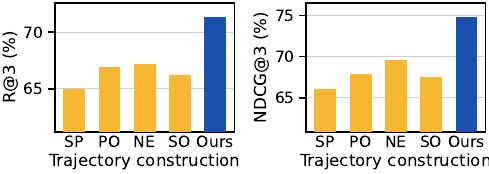}
\caption{Impact of trajectory construction.}
\label{fig:trajectory-construction}
\end{minipage}
\end{figure}

To explore how the scale of the backbone affects ReMind, we initialize ReMind from Qwen3-Reranker-0.6B, 4B, and 8B, and evaluate it on PersonaBench with the dense retriever.
Figure~\ref{fig:backbone-scale} shows R@3 and NDCG@3, and Table~\ref{tab:backbone-scale} reports all metrics.
We can draw two conclusions.
First, a larger backbone brings better performance, and the trend holds on every metric.
Second, the improvement from 0.6B to 4B is much larger than that from 4B to 8B: R@3 rises from 31.28 to 71.30 and NDCG@3 from 27.57 to 74.77 between 0.6B and 4B, while the two gains from 4B to 8B are 4.56 and 3.23.
We attribute the weak result of 0.6B to its small capacity, which is insufficient for recurrent ranking, where the model must judge relevance conditioned on a growing context and predict sufficiency at the same time.
It even falls below the original Qwen3-Reranker-0.6B in Table~\ref{tab:main-dense}.
The small gap between 4B and 8B indicates that the 4B backbone already reaches a good balance between effectiveness and the cost of training and deployment, so we use it as the default.

\begin{table}[h]
\caption{Complete results of ReMind with different backbone scales on PersonaBench with dense retrieval. Bold marks the best result in each column and underline marks the second best.}
\label{tab:backbone-scale}
\small
\centering
\setlength{\tabcolsep}{3pt}
\renewcommand{\arraystretch}{1.0}
\begin{tabular*}{\textwidth}{@{\hspace{2pt}\extracolsep{\fill}}l*{9}{c}@{\hspace{2pt}}}
\toprule
\textbf{Backbone} & \textbf{R@1} & \textbf{R@3} & \textbf{R@5} & \textbf{N@1} & \textbf{N@3} & \textbf{N@5} & \textbf{MAP} & \textbf{MRR} & \textbf{F1} \\
\midrule
0.6B & 7.61 & 31.28 & 40.97 & 17.28 & 27.57 & 30.92 & 29.48 & 34.55 & 21.97 \\
4B & \underline{49.81} & \underline{71.30} & \underline{78.74} & \underline{72.84} & \underline{74.77} & \underline{74.28} & \underline{72.61} & \underline{79.25} & \underline{48.09} \\
8B & \textbf{50.43} & \textbf{75.86} & \textbf{83.31} & \textbf{74.07} & \textbf{78.00} & \textbf{77.42} & \textbf{74.55} & \textbf{82.00} & \textbf{50.74} \\
\bottomrule
\end{tabular*}
\end{table}

\section{Training Trajectory Construction}
\label{app:trajectory-construction}
We study how the construction of training trajectories affects ReMind on PersonaBench with the dense retriever.
Figure~\ref{fig:trajectory-construction} reports R@3 and NDCG@3, and Table~\ref{tab:trajectory-construction} reports all metrics.
We compare ReMind with the following variants, each of which changes one choice in the trajectory construction:
\begin{itemize}[leftmargin=*,nosep]
    \item \textbf{Single-pass (SP)}: no trajectories are built; each query is paired with one positive memory and the hard negatives, and the model is trained without the selected-memory context.
    \item \textbf{Positive-only (PO)}: trajectories contain only positive memories, so the context never includes a negative memory ($\rho=0$).
    \item \textbf{Negative-end (NE)}: negative memories are also used as selection targets, so a trajectory may end with a negative memory.
    \item \textbf{Single-order (SO)}: only one ordering of the positive memories is sampled for each query ($P=1$).
\end{itemize}
We can draw the following conclusions.
First, our trajectory construction outperforms all the variants on both metrics, which shows that every design choice contributes to the final performance.
Second, compared with SP and PO, ReMind trains on recurrent trajectories whose context mixes positive and negative memories: the gain over SP shows the value of training on selection states rather than on isolated query-memory pairs, and the gain over PO, 4.35 on R@3 and 6.91 on NDCG@3, shows that these states must contain negative memories, as they do at inference.
Third, NE is the strongest variant, so exposing the ranker to negative memories helps, but supervising it to select them injects noise into the training signal and still costs 4.08 on R@3 and 5.23 on NDCG@3 compared with ReMind, which uses negative memories only in the context and as candidates to reject.
Fourth, ReMind outperforms SO by 5.07 on R@3 and 7.24 on NDCG@3, which shows that a set of positive memories can yield more than one useful trajectory: different orderings of the same positives serve as data augmentation and expose the ranker to more selection states.

\begin{table}[h]
\caption{Complete results of different training trajectory constructions on PersonaBench with dense retrieval. Bold marks the best result in each column and underline marks the second best.}
\label{tab:trajectory-construction}
\small
\centering
\setlength{\tabcolsep}{3pt}
\renewcommand{\arraystretch}{1.0}
\begin{tabular*}{\textwidth}{@{\hspace{2pt}\extracolsep{\fill}}l*{9}{c}@{\hspace{2pt}}}
\toprule
\textbf{Variant} & \textbf{R@1} & \textbf{R@3} & \textbf{R@5} & \textbf{N@1} & \textbf{N@3} & \textbf{N@5} & \textbf{MAP} & \textbf{MRR} & \textbf{F1} \\
\midrule
SP & 41.77 & 65.00 & 75.84 & 59.26 & 66.08 & 67.84 & 65.50 & 70.83 & 43.58 \\
PO & 41.93 & 66.95 & 76.03 & 60.49 & 67.86 & 68.74 & 65.98 & 72.10 & 45.33 \\
NE & \underline{45.68} & \underline{67.22} & \underline{77.20} & \underline{65.43} & \underline{69.54} & \underline{70.82} & \underline{68.67} & \underline{74.67} & \underline{45.49} \\
SO & 40.23 & 66.23 & 75.99 & 59.26 & 67.53 & 68.71 & 66.20 & 72.33 & 44.78 \\
\midrule
\textbf{ReMind} & \textbf{49.81} & \textbf{71.30} & \textbf{78.74} & \textbf{72.84} & \textbf{74.77} & \textbf{74.28} & \textbf{72.61} & \textbf{79.25} & \textbf{48.09} \\
\bottomrule
\end{tabular*}
\end{table}

\section{Additional Information about training and inference}
\label{app:key-parameters}

In this section, we provide more details on training and inference, as shown in Table~\ref{tab:more_hyper}.

\begin{table}
\centering
\caption{More Details on Training and Inference.}
\label{tab:more_hyper}
\resizebox{0.8\linewidth}{!}{
\begin{tabular}{cc} 
\toprule
Parameter            & Setting                                                \\ 
\midrule
GPU                  & NVIDIA RTX 4090                                  \\
Training Time        & 2.5 hours / epoch  (PersonaBench)                      \\
Total params         &  4.0548 B                                      \\
Trainable params     &  33.0301M (0.8146\%)                            \\
Batch Size           & 16                                             \\
Training GPU Memory Consumption           &   8.46-10.39 GB / GPU                                             \\
Backbone       & Huggingface Qwen3-Reranker-4B          \\
Optimizer & AdamW~\citep{loshchilov2017decoupled}\\
Epoch & 1epoch\\
Learning rate & $2\times10^{-4}$ \\
Warmup ratio & 0.1 \\
Weight decay & 0.01 \\
\bottomrule
\end{tabular}
}
\end{table}

\section{Complete Main Results}
\label{app:complete-main-results}
Table~\ref{tab:main-dense} in the main text reports the results with the dense retriever on LongMemEval-M, PersonaBench, and ConvoMem.
This section reports the complete main results.
Tables~\ref{tab:main-complete-lme-s}, \ref{tab:main-complete-lme-m}, \ref{tab:main-complete-pb}, and \ref{tab:main-complete-cm} list the results of all methods on LongMemEval-S, LongMemEval-M, PersonaBench, and ConvoMem, respectively, each with the BM25, dense, and RF-Mem retrievers; the dense blocks of the last three tables repeat Table~\ref{tab:main-dense}.

\begin{table}[h]
\caption{Complete main results on LongMemEval-S with the BM25, dense, and RF-Mem retrievers (\%). Within each retriever, bold marks the best result in each column and underline marks the second best.}
\label{tab:main-complete-lme-s}
\small
\centering
\setlength{\tabcolsep}{2pt}
\renewcommand{\arraystretch}{0.95}
\begin{tabular*}{\textwidth}{@{\hspace{2pt}\extracolsep{\fill}}l*{9}{c}@{\hspace{2pt}}}
\toprule
\textbf{Method} & \textbf{R@1} & \textbf{R@3} & \textbf{R@5} & \textbf{N@1} & \textbf{N@3} & \textbf{N@5} & \textbf{MAP} & \textbf{MRR} & \textbf{F1} \\
\midrule
\multicolumn{10}{@{}l@{}}{\makebox[0pt][l]{\colorbox[gray]{0.95}{\makebox[\dimexpr\textwidth-2\fboxsep\relax]{\strut\textbf{\textit{BM25}}}}}} \\
Base & 50.20 & 84.04 & 90.23 & 84.21 & 84.67 & 86.67 & 84.93 & 89.38 & 62.25 \\
\midrule
Qwen3-Reranker-0.6B~\citep{zhang2025qwen3} & 48.14 & 75.89 & 86.44 & 76.02 & 76.46 & 81.16 & 79.76 & 84.14 & 55.32 \\
Qwen3-Reranker-4B~\citep{zhang2025qwen3} & 46.08 & 79.89 & 93.97 & 75.44 & 78.85 & 84.76 & 81.38 & 84.40 & 58.76 \\
Qwen3-Reranker-8B~\citep{zhang2025qwen3} & 56.71 & 89.12 & 96.50 & 92.40 & 91.01 & 93.43 & 91.76 & 95.26 & 65.71 \\
KaLM-Nano~\citep{zhao2026kalm} & 52.11 & 83.26 & 93.33 & 87.13 & 84.94 & 88.80 & 86.88 & 91.05 & 61.70 \\
KaLM-Small~\citep{zhao2026kalm} & 55.83 & 91.36 & 96.72 & 92.40 & 92.95 & 94.26 & 93.22 & 95.51 & 67.95 \\
KaLM-Large~\citep{zhao2026kalm} & 53.98 & 89.20 & 97.02 & 90.06 & 90.37 & 93.01 & 91.44 & 93.90 & 66.11 \\
RankGPT~\citep{sun2023chatgpt} & 53.15 & 92.37 & 95.56 & 90.06 & 92.82 & 93.10 & 92.16 & 93.82 & 69.07 \\
\midrule
MemReranker-0.6B~\citep{li2026memreranker} & 57.10 & 94.82 & \underline{99.76} & 94.74 & 96.20 & 97.45 & 96.30 & 97.08 & 70.67 \\
MemReranker-4B~\citep{li2026memreranker} & \textbf{61.19} & \underline{96.24} & \textbf{99.90} & \textbf{100.00} & \underline{99.09} & \underline{99.66} & \underline{99.29} & \textbf{100.00} & \underline{71.93} \\
QRRanker~\citep{li2026query} & 53.68 & 89.44 & 93.66 & 90.06 & 90.36 & 91.27 & 90.14 & 93.30 & 66.86 \\
ConvMemory~\citep{pan2026convmemory,pan2026convmemoryv2} & 47.45 & 84.00 & 90.09 & 80.70 & 83.29 & 85.35 & 83.60 & 87.23 & 62.40 \\
\midrule
\textbf{ReMind} & \underline{60.60} & \textbf{97.31} & \textbf{99.90} & \underline{99.42} & \textbf{99.69} & \textbf{99.69} & \textbf{99.51} & \underline{99.71} & \textbf{72.83} \\
\midrule
\multicolumn{10}{@{}l@{}}{\makebox[0pt][l]{\colorbox[gray]{0.95}{\makebox[\dimexpr\textwidth-2\fboxsep\relax]{\strut\textbf{\textit{Dense retriever}}}}}} \\
Base & 38.19 & 70.09 & 79.85 & 68.42 & 69.27 & 72.92 & 70.36 & 77.61 & 52.61 \\
\midrule
Qwen3-Reranker-0.6B~\citep{zhang2025qwen3} & 47.85 & 76.11 & 85.96 & 75.44 & 76.55 & 80.84 & 79.64 & 83.80 & 55.54 \\
Qwen3-Reranker-4B~\citep{zhang2025qwen3} & 43.87 & 69.99 & 90.12 & 70.76 & 70.22 & 79.25 & 75.88 & 79.79 & 50.85 \\
Qwen3-Reranker-8B~\citep{zhang2025qwen3} & 54.93 & 82.53 & 93.79 & 88.89 & 84.95 & 89.52 & 87.70 & 92.37 & 60.46 \\
KaLM-Nano~\citep{zhao2026kalm} & 50.45 & 79.48 & 90.53 & 84.80 & 81.26 & 85.55 & 83.66 & 89.13 & 58.64 \\
KaLM-Small~\citep{zhao2026kalm} & 55.83 & 88.88 & 95.84 & 92.40 & 91.15 & 93.27 & 92.02 & 95.35 & 65.97 \\
KaLM-Large~\citep{zhao2026kalm} & 52.66 & 86.03 & 94.12 & 88.30 & 87.45 & 90.16 & 88.92 & 92.41 & 63.70 \\
RankGPT~\citep{sun2023chatgpt} & 54.46 & 92.22 & 97.08 & 91.81 & 93.13 & 94.18 & 92.60 & 95.14 & 69.02 \\
\midrule
MemReranker-0.6B~\citep{li2026memreranker} & 56.80 & 94.82 & 99.46 & 94.15 & 96.11 & 97.23 & 96.18 & 96.88 & 70.67 \\
MemReranker-4B~\citep{li2026memreranker} & \underline{60.46} & \underline{95.51} & \textbf{99.76} & \underline{98.83} & \underline{98.07} & \underline{98.92} & \underline{98.74} & \underline{99.27} & \underline{71.47} \\
QRRanker~\citep{li2026query} & 51.18 & 89.79 & 96.38 & 86.55 & 89.80 & 92.05 & 90.02 & 92.08 & 67.12 \\
ConvMemory~\citep{pan2026convmemory,pan2026convmemoryv2} & 46.47 & 83.06 & 92.24 & 80.12 & 82.33 & 85.80 & 83.18 & 87.19 & 61.66 \\
\midrule
\textbf{ReMind-SFT} & 58.12 & 95.29 & 98.91 & 95.91 & 97.32 & 97.96 & 97.11 & 97.51 & 71.17 \\
\textbf{ReMind} & \textbf{60.90} & \textbf{96.73} & \underline{99.61} & \textbf{99.42} & \textbf{99.32} & \textbf{99.48} & \textbf{99.39} & \textbf{99.56} & \textbf{72.36} \\
\midrule
\multicolumn{10}{@{}l@{}}{\makebox[0pt][l]{\colorbox[gray]{0.95}{\makebox[\dimexpr\textwidth-2\fboxsep\relax]{\strut\textbf{\textit{RF-Mem}}}}}} \\
Base & 38.19 & 70.09 & 79.58 & 68.42 & 69.27 & 72.97 & 70.67 & 77.46 & 52.61 \\
\midrule
Qwen3-Reranker-0.6B~\citep{zhang2025qwen3} & 50.53 & 79.33 & 89.34 & 80.70 & 80.34 & 84.44 & 82.67 & 87.05 & 58.36 \\
Qwen3-Reranker-4B~\citep{zhang2025qwen3} & 43.77 & 69.89 & 89.82 & 70.76 & 70.10 & 79.05 & 75.79 & 79.65 & 50.81 \\
Qwen3-Reranker-8B~\citep{zhang2025qwen3} & 54.64 & 82.24 & 93.67 & 88.30 & 84.70 & 89.36 & 87.61 & 91.93 & 60.22 \\
KaLM-Nano~\citep{zhao2026kalm} & 50.45 & 79.39 & 90.23 & 84.80 & 81.22 & 85.39 & 83.64 & 89.03 & 58.60 \\
KaLM-Small~\citep{zhao2026kalm} & 55.73 & 88.88 & 95.55 & 92.40 & 91.11 & 93.07 & 91.89 & 95.25 & 65.97 \\
KaLM-Large~\citep{zhao2026kalm} & 52.37 & 86.23 & 94.32 & 87.72 & 87.45 & 90.15 & 88.76 & 92.05 & 63.89 \\
RankGPT~\citep{sun2023chatgpt} & 54.76 & 92.32 & 95.89 & 92.40 & 93.15 & 93.63 & 92.40 & 95.35 & 68.88 \\
\midrule
MemReranker-0.6B~\citep{li2026memreranker} & 57.53 & 95.02 & 99.17 & \underline{94.74} & 96.61 & 97.47 & 96.62 & \underline{97.27} & 70.87 \\
MemReranker-4B~\citep{li2026memreranker} & \underline{60.60} & \underline{96.14} & \textbf{99.61} & \textbf{99.42} & \underline{98.40} & \underline{98.82} & \underline{98.65} & \textbf{99.61} & \underline{71.89} \\
QRRanker~\citep{li2026query} & 52.56 & 90.84 & 94.66 & 88.89 & 91.11 & 91.95 & 90.87 & 93.15 & 67.85 \\
ConvMemory~\citep{pan2026convmemory,pan2026convmemoryv2} & 45.89 & 84.72 & 92.83 & 79.53 & 83.24 & 86.15 & 83.49 & 87.19 & 62.85 \\
\midrule
\textbf{ReMind} & \textbf{60.90} & \textbf{96.73} & \underline{99.32} & \textbf{99.42} & \textbf{99.32} & \textbf{99.32} & \textbf{99.22} & \textbf{99.61} & \textbf{72.36} \\
\bottomrule
\end{tabular*}
\end{table}

\begin{table}[h]
\caption{Complete main results on LongMemEval-M with the BM25, dense, and RF-Mem retrievers (\%). Within each retriever, bold marks the best result in each column and underline marks the second best.}
\label{tab:main-complete-lme-m}
\small
\centering
\setlength{\tabcolsep}{2pt}
\renewcommand{\arraystretch}{0.95}
\begin{tabular*}{\textwidth}{@{\hspace{2pt}\extracolsep{\fill}}l*{9}{c}@{\hspace{2pt}}}
\toprule
\textbf{Method} & \textbf{R@1} & \textbf{R@3} & \textbf{R@5} & \textbf{N@1} & \textbf{N@3} & \textbf{N@5} & \textbf{MAP} & \textbf{MRR} & \textbf{F1} \\
\midrule
\multicolumn{10}{@{}l@{}}{\makebox[0pt][l]{\colorbox[gray]{0.95}{\makebox[\dimexpr\textwidth-2\fboxsep\relax]{\strut\textbf{\textit{BM25}}}}}} \\
Base & 39.51 & 67.17 & 74.34 & 67.46 & 66.33 & 69.45 & 67.17 & 75.27 & 49.68 \\
\midrule
Qwen3-Reranker-0.6B~\citep{zhang2025qwen3} & 40.62 & 69.42 & 78.20 & 67.46 & 68.58 & 72.39 & 69.77 & 77.66 & 50.82 \\
Qwen3-Reranker-4B~\citep{zhang2025qwen3} & 51.87 & 86.12 & 90.63 & 87.57 & 86.97 & 88.12 & 85.71 & 91.52 & 64.53 \\
Qwen3-Reranker-8B~\citep{zhang2025qwen3} & 51.68 & 87.16 & 91.88 & 88.17 & 87.96 & 89.37 & 86.87 & 92.45 & 65.03 \\
KaLM-Nano~\citep{zhao2026kalm} & 47.14 & 79.18 & 86.52 & 81.66 & 79.82 & 82.47 & 79.73 & 86.97 & 59.20 \\
KaLM-Small~\citep{zhao2026kalm} & \underline{53.70} & 86.42 & 90.78 & 89.94 & 88.14 & 89.41 & 87.54 & 92.95 & 64.56 \\
KaLM-Large~\citep{zhao2026kalm} & 52.49 & 87.24 & 92.16 & 88.17 & 88.19 & 89.72 & 87.34 & 92.46 & 64.96 \\
RankGPT~\citep{sun2023chatgpt} & 43.63 & 81.10 & 85.49 & 77.51 & 80.11 & 81.28 & 78.97 & 84.86 & 61.28 \\
\midrule
MemReranker-0.6B~\citep{li2026memreranker} & 50.94 & 86.73 & 90.65 & 86.39 & 87.05 & 88.14 & 85.96 & 90.94 & 64.99 \\
MemReranker-4B~\citep{li2026memreranker} & \textbf{54.73} & \underline{90.35} & \underline{92.63} & \textbf{93.49} & \underline{91.63} & \underline{91.53} & \underline{90.82} & \textbf{95.22} & \underline{68.17} \\
QRRanker~\citep{li2026query} & 43.96 & 75.37 & 82.33 & 75.74 & 75.00 & 77.74 & 75.22 & 82.44 & 55.76 \\
ConvMemory~\citep{pan2026convmemory,pan2026convmemoryv2} & 32.92 & 63.85 & 75.62 & 57.99 & 61.20 & 66.45 & 62.62 & 71.04 & 46.85 \\
\midrule
\textbf{ReMind} & 53.18 & \textbf{90.53} & \textbf{93.55} & \underline{91.72} & \textbf{92.24} & \textbf{92.63} & \textbf{91.23} & \underline{94.69} & \textbf{68.62} \\
\midrule
\multicolumn{10}{@{}l@{}}{\makebox[0pt][l]{\colorbox[gray]{0.95}{\makebox[\dimexpr\textwidth-2\fboxsep\relax]{\strut\textbf{\textit{Dense retriever}}}}}} \\
Base & 24.62 & 46.40 & 53.83 & 43.79 & 44.24 & 47.71 & 45.80 & 54.64 & 33.74 \\
\midrule
Qwen3-Reranker-0.6B~\citep{zhang2025qwen3} & 39.73 & 64.88 & 70.52 & 68.64 & 64.77 & 67.19 & 64.15 & 75.93 & 47.50 \\
Qwen3-Reranker-4B~\citep{zhang2025qwen3} & 45.08 & 73.14 & 79.89 & 75.74 & 74.19 & 76.50 & 72.99 & 81.65 & 55.06 \\
Qwen3-Reranker-8B~\citep{zhang2025qwen3} & 44.85 & 73.42 & 80.83 & 76.33 & 74.45 & 77.35 & 73.72 & 82.18 & 54.74 \\
KaLM-Nano~\citep{zhao2026kalm} & 39.69 & 66.77 & 74.96 & 70.41 & 67.56 & 70.84 & 67.34 & 76.91 & 50.04 \\
KaLM-Small~\citep{zhao2026kalm} & 48.32 & 75.90 & 79.34 & 84.02 & 78.49 & 79.24 & 76.73 & 86.39 & 57.29 \\
KaLM-Large~\citep{zhao2026kalm} & 48.18 & 75.70 & 81.39 & 82.84 & 78.01 & 79.81 & 76.77 & 86.08 & 56.87 \\
RankGPT~\citep{sun2023chatgpt} & 43.72 & 70.70 & 73.75 & 76.92 & 72.23 & 72.67 & 70.42 & 80.69 & 53.22 \\
\midrule
MemReranker-0.6B~\citep{li2026memreranker} & 46.08 & 76.12 & 80.93 & 79.88 & 77.66 & 79.11 & 76.13 & 84.49 & 57.57 \\
MemReranker-4B~\citep{li2026memreranker} & \underline{49.21} & \underline{79.31} & \underline{82.41} & \underline{85.80} & \underline{81.81} & \underline{82.14} & \underline{79.57} & \underline{87.70} & \underline{60.18} \\
QRRanker~\citep{li2026query} & 35.65 & 63.63 & 70.78 & 64.50 & 63.29 & 66.02 & 62.74 & 73.07 & 47.73 \\
ConvMemory~\citep{pan2026convmemory,pan2026convmemoryv2} & 34.95 & 59.44 & 69.23 & 60.36 & 58.77 & 63.30 & 59.53 & 69.54 & 43.89 \\
\midrule
\textbf{ReMind-SFT} & 49.03 & 77.35 & 80.79 & 85.21 & 79.99 & 80.63 & 77.95 & 87.31 & 58.37 \\
\textbf{ReMind} & \textbf{50.07} & \textbf{80.53} & \textbf{82.82} & \textbf{86.39} & \textbf{83.17} & \textbf{83.03} & \textbf{80.74} & \textbf{88.41} & \textbf{60.87} \\
\midrule
\multicolumn{10}{@{}l@{}}{\makebox[0pt][l]{\colorbox[gray]{0.95}{\makebox[\dimexpr\textwidth-2\fboxsep\relax]{\strut\textbf{\textit{RF-Mem}}}}}} \\
Base & 24.62 & 46.40 & 54.65 & 43.79 & 44.24 & 48.19 & 46.24 & 54.79 & 33.74 \\
\midrule
Qwen3-Reranker-0.6B~\citep{zhang2025qwen3} & 40.87 & 66.31 & 73.28 & 69.23 & 66.34 & 69.32 & 66.23 & 76.77 & 48.83 \\
Qwen3-Reranker-4B~\citep{zhang2025qwen3} & 47.56 & 74.50 & 82.23 & 79.88 & 76.24 & 79.16 & 75.87 & 84.48 & 55.74 \\
Qwen3-Reranker-8B~\citep{zhang2025qwen3} & 47.07 & 75.43 & 82.89 & 79.29 & 76.85 & 79.74 & 76.49 & 84.40 & 56.18 \\
KaLM-Nano~\citep{zhao2026kalm} & 42.21 & 68.39 & 77.00 & 73.96 & 69.64 & 73.20 & 69.97 & 79.30 & 51.11 \\
KaLM-Small~\citep{zhao2026kalm} & 49.36 & 78.51 & 82.37 & 85.21 & 80.63 & 81.57 & 79.15 & 87.68 & 59.11 \\
KaLM-Large~\citep{zhao2026kalm} & 49.51 & 78.91 & 84.02 & 84.62 & 80.76 & 82.42 & 79.67 & 87.97 & 58.99 \\
RankGPT~\citep{sun2023chatgpt} & 43.32 & 72.28 & 75.44 & 77.51 & 73.62 & 74.06 & 71.51 & 82.10 & 54.92 \\
\midrule
MemReranker-0.6B~\citep{li2026memreranker} & 48.01 & 77.85 & 83.24 & 82.25 & 79.61 & 81.39 & 78.73 & 86.38 & 58.68 \\
MemReranker-4B~\citep{li2026memreranker} & \underline{51.13} & \underline{81.75} & \underline{84.96} & \underline{88.17} & \underline{84.34} & \underline{84.77} & \underline{82.44} & \underline{89.77} & \underline{61.91} \\
QRRanker~\citep{li2026query} & 41.15 & 67.28 & 73.92 & 71.01 & 67.85 & 70.61 & 67.54 & 78.30 & 49.78 \\
ConvMemory~\citep{pan2026convmemory,pan2026convmemoryv2} & 34.99 & 60.43 & 70.06 & 60.95 & 59.75 & 64.20 & 60.48 & 70.57 & 44.72 \\
\midrule
\textbf{ReMind} & \textbf{52.14} & \textbf{82.85} & \textbf{85.37} & \textbf{89.35} & \textbf{85.67} & \textbf{85.74} & \textbf{83.68} & \textbf{90.78} & \textbf{62.46} \\
\bottomrule
\end{tabular*}
\end{table}

\begin{table}[h]
\caption{Complete main results on PersonaBench with the BM25, dense, and RF-Mem retrievers (\%). Within each retriever, bold marks the best result in each column and underline marks the second best.}
\label{tab:main-complete-pb}
\small
\centering
\setlength{\tabcolsep}{2pt}
\renewcommand{\arraystretch}{0.95}
\begin{tabular*}{\textwidth}{@{\hspace{2pt}\extracolsep{\fill}}l*{9}{c}@{\hspace{2pt}}}
\toprule
\textbf{Method} & \textbf{R@1} & \textbf{R@3} & \textbf{R@5} & \textbf{N@1} & \textbf{N@3} & \textbf{N@5} & \textbf{MAP} & \textbf{MRR} & \textbf{F1} \\
\midrule
\multicolumn{10}{@{}l@{}}{\makebox[0pt][l]{\colorbox[gray]{0.95}{\makebox[\dimexpr\textwidth-2\fboxsep\relax]{\strut\textbf{\textit{BM25}}}}}} \\
Base & 5.02 & 14.47 & 19.65 & 8.64 & 13.08 & 14.65 & 15.76 & 19.55 & 10.74 \\
\midrule
Qwen3-Reranker-0.6B~\citep{zhang2025qwen3} & 19.47 & 33.02 & 42.86 & 33.33 & 34.12 & 36.55 & 34.38 & 42.42 & 23.25 \\
Qwen3-Reranker-4B~\citep{zhang2025qwen3} & 31.32 & \underline{49.20} & 55.76 & 48.15 & 50.99 & 51.62 & 48.61 & 56.35 & 33.25 \\
Qwen3-Reranker-8B~\citep{zhang2025qwen3} & 34.88 & 46.98 & 53.27 & 50.62 & 51.35 & 51.56 & \underline{49.91} & 57.58 & 32.32 \\
KaLM-Nano~\citep{zhao2026kalm} & 19.61 & 39.38 & 47.35 & 34.57 & 39.01 & 40.40 & 37.20 & 45.34 & 26.81 \\
KaLM-Small~\citep{zhao2026kalm} & 28.79 & 49.12 & 55.74 & 48.15 & 49.66 & 50.08 & 46.78 & 56.41 & 32.79 \\
KaLM-Large~\citep{zhao2026kalm} & 31.87 & 47.98 & \underline{56.85} & 49.38 & \underline{51.37} & \underline{52.29} & 49.37 & 56.72 & \underline{33.74} \\
RankGPT~\citep{sun2023chatgpt} & 28.54 & 41.44 & 43.25 & 46.91 & 44.78 & 42.77 & 41.60 & 52.92 & 28.04 \\
\midrule
MemReranker-0.6B~\citep{li2026memreranker} & 25.88 & 43.66 & 51.30 & 40.74 & 45.05 & 46.06 & 43.56 & 50.84 & 30.47 \\
MemReranker-4B~\citep{li2026memreranker} & \underline{35.33} & 45.21 & 52.33 & \underline{53.09} & 51.22 & 51.14 & 49.87 & \underline{58.61} & 31.78 \\
QRRanker~\citep{li2026query} & 28.19 & 42.84 & 54.40 & 38.27 & 43.39 & 47.20 & 44.06 & 49.62 & 28.14 \\
ConvMemory~\citep{pan2026convmemory,pan2026convmemoryv2} & 16.67 & 27.24 & 33.99 & 25.93 & 28.50 & 29.38 & 29.52 & 35.73 & 19.21 \\
\midrule
\textbf{ReMind} & \textbf{37.67} & \textbf{53.07} & \textbf{60.60} & \textbf{58.02} & \textbf{57.45} & \textbf{57.53} & \textbf{55.03} & \textbf{63.09} & \textbf{36.50} \\
\midrule
\multicolumn{10}{@{}l@{}}{\makebox[0pt][l]{\colorbox[gray]{0.95}{\makebox[\dimexpr\textwidth-2\fboxsep\relax]{\strut\textbf{\textit{Dense retriever}}}}}} \\
Base & 17.41 & 33.07 & 42.47 & 30.86 & 33.28 & 35.90 & 36.52 & 43.54 & 22.67 \\
\midrule
Qwen3-Reranker-0.6B~\citep{zhang2025qwen3} & 24.47 & 38.91 & 53.27 & 38.27 & 39.98 & 44.64 & 43.14 & 51.25 & 26.17 \\
Qwen3-Reranker-4B~\citep{zhang2025qwen3} & 35.70 & 60.78 & 70.37 & 54.32 & 60.41 & 62.64 & 59.39 & 65.88 & 40.37 \\
Qwen3-Reranker-8B~\citep{zhang2025qwen3} & 39.20 & 60.04 & 69.96 & 56.79 & 61.20 & 63.01 & 61.03 & 66.94 & 40.37 \\
KaLM-Nano~\citep{zhao2026kalm} & 22.45 & 46.54 & 58.17 & 38.27 & 44.90 & 47.88 & 45.55 & 52.64 & 31.31 \\
KaLM-Small~\citep{zhao2026kalm} & 31.38 & 58.37 & 69.07 & 50.62 & 56.47 & 58.70 & 55.52 & 63.17 & 38.35 \\
KaLM-Large~\citep{zhao2026kalm} & 37.67 & 54.63 & 70.35 & 58.02 & 58.57 & 62.30 & 59.65 & 66.46 & 38.02 \\
RankGPT~\citep{sun2023chatgpt} & 33.79 & 52.04 & 61.03 & 53.09 & 54.25 & 54.81 & 53.95 & 62.02 & 35.99 \\
\midrule
MemReranker-0.6B~\citep{li2026memreranker} & 31.19 & 49.96 & 64.01 & 46.91 & 51.03 & 55.13 & 53.24 & 58.47 & 34.17 \\
MemReranker-4B~\citep{li2026memreranker} & 33.48 & 51.30 & 61.73 & 53.09 & 54.35 & 56.09 & 55.47 & 62.22 & 35.78 \\
QRRanker~\citep{li2026query} & 26.07 & 50.80 & 65.08 & 44.44 & 49.42 & 53.64 & 50.44 & 58.96 & 33.62 \\
ConvMemory~\citep{pan2026convmemory,pan2026convmemoryv2} & 17.59 & 32.41 & 40.88 & 27.16 & 32.60 & 34.52 & 35.26 & 40.24 & 22.91 \\
\midrule
\textbf{ReMind-SFT} & \underline{41.77} & \underline{65.00} & \underline{75.84} & \underline{59.26} & \underline{66.08} & \underline{67.84} & \underline{65.50} & \underline{70.83} & \underline{43.58} \\
\textbf{ReMind} & \textbf{49.81} & \textbf{71.30} & \textbf{78.74} & \textbf{72.84} & \textbf{74.77} & \textbf{74.28} & \textbf{72.61} & \textbf{79.25} & \textbf{48.09} \\
\midrule
\multicolumn{10}{@{}l@{}}{\makebox[0pt][l]{\colorbox[gray]{0.95}{\makebox[\dimexpr\textwidth-2\fboxsep\relax]{\strut\textbf{\textit{RF-Mem}}}}}} \\
Base & 12.47 & 26.52 & 42.72 & 25.93 & 27.54 & 32.79 & 31.87 & 39.24 & 19.40 \\
\midrule
Qwen3-Reranker-0.6B~\citep{zhang2025qwen3} & 23.23 & 37.30 & 54.51 & 37.04 & 38.63 & 44.23 & 41.57 & 49.61 & 25.73 \\
Qwen3-Reranker-4B~\citep{zhang2025qwen3} & 34.47 & \underline{56.71} & \underline{66.50} & 53.09 & 57.52 & \underline{59.72} & 56.21 & 63.61 & \underline{38.33} \\
Qwen3-Reranker-8B~\citep{zhang2025qwen3} & \underline{37.96} & 56.54 & 63.74 & \underline{55.56} & \underline{58.75} & 59.37 & \underline{58.11} & \underline{65.07} & 38.23 \\
KaLM-Nano~\citep{zhao2026kalm} & 23.93 & 46.17 & 55.53 & 40.74 & 45.64 & 47.47 & 45.37 & 53.55 & 31.13 \\
KaLM-Small~\citep{zhao2026kalm} & 31.38 & 54.05 & 64.34 & 50.62 & 53.98 & 55.98 & 53.09 & 61.61 & 36.00 \\
KaLM-Large~\citep{zhao2026kalm} & 35.21 & 52.37 & 64.38 & \underline{55.56} & 55.79 & 57.90 & 55.68 & 63.45 & 36.50 \\
RankGPT~\citep{sun2023chatgpt} & 35.02 & 51.05 & 57.20 & 54.32 & 54.45 & 53.87 & 53.59 & 61.73 & 35.68 \\
\midrule
MemReranker-0.6B~\citep{li2026memreranker} & 31.44 & 47.94 & 59.53 & 48.15 & 50.16 & 53.09 & 51.71 & 57.97 & 32.96 \\
MemReranker-4B~\citep{li2026memreranker} & 32.24 & 49.28 & 58.23 & 51.85 & 52.96 & 53.96 & 53.39 & 61.03 & 34.56 \\
QRRanker~\citep{li2026query} & 30.19 & 51.42 & 62.59 & 49.38 & 51.39 & 53.86 & 51.37 & 61.22 & 33.74 \\
ConvMemory~\citep{pan2026convmemory,pan2026convmemoryv2} & 19.49 & 32.72 & 42.61 & 30.86 & 33.85 & 36.30 & 36.39 & 42.02 & 23.26 \\
\midrule
\textbf{ReMind} & \textbf{42.98} & \textbf{63.42} & \textbf{73.33} & \textbf{64.20} & \textbf{66.50} & \textbf{67.73} & \textbf{65.30} & \textbf{72.34} & \textbf{43.21} \\
\bottomrule
\end{tabular*}
\end{table}

\begin{table}[h]
\caption{Complete main results on ConvoMem with the BM25, dense, and RF-Mem retrievers (\%). Within each retriever, bold marks the best result in each column and underline marks the second best.}
\label{tab:main-complete-cm}
\small
\centering
\setlength{\tabcolsep}{2pt}
\renewcommand{\arraystretch}{0.95}
\begin{tabular*}{\textwidth}{@{\hspace{2pt}\extracolsep{\fill}}l*{9}{c}@{\hspace{2pt}}}
\toprule
\textbf{Method} & \textbf{R@1} & \textbf{R@3} & \textbf{R@5} & \textbf{N@1} & \textbf{N@3} & \textbf{N@5} & \textbf{MAP} & \textbf{MRR} & \textbf{F1} \\
\midrule
\multicolumn{10}{@{}l@{}}{\makebox[0pt][l]{\colorbox[gray]{0.95}{\makebox[\dimexpr\textwidth-2\fboxsep\relax]{\strut\textbf{\textit{BM25}}}}}} \\
Base & 17.22 & 37.89 & 46.59 & 51.50 & 46.00 & 46.66 & 45.32 & 60.83 & 37.13 \\
\midrule
Qwen3-Reranker-0.6B~\citep{zhang2025qwen3} & 21.23 & 49.17 & 58.55 & 68.50 & 62.12 & 60.60 & 57.26 & 76.04 & 49.99 \\
Qwen3-Reranker-4B~\citep{zhang2025qwen3} & 24.00 & 57.75 & 68.94 & 78.50 & 72.76 & 70.98 & 67.06 & 82.98 & 59.31 \\
Qwen3-Reranker-8B~\citep{zhang2025qwen3} & 22.99 & 54.88 & 66.08 & 75.00 & 69.59 & 68.21 & 64.42 & 81.42 & 56.32 \\
KaLM-Nano~\citep{zhao2026kalm} & 24.67 & 55.77 & 65.69 & 78.50 & 70.61 & 68.56 & 65.23 & 82.57 & 56.82 \\
KaLM-Small~\citep{zhao2026kalm} & 25.60 & 58.38 & 68.72 & \underline{81.50} & 73.52 & 71.67 & 67.84 & 84.98 & 59.39 \\
KaLM-Large~\citep{zhao2026kalm} & \underline{25.70} & \underline{59.73} & \underline{70.30} & \underline{81.50} & \underline{75.06} & \underline{73.12} & \underline{69.00} & \underline{85.27} & \underline{60.91} \\
RankGPT~\citep{sun2023chatgpt} & 22.66 & 53.43 & 60.73 & 72.50 & 66.87 & 63.69 & 60.88 & 79.38 & 54.34 \\
\midrule
MemReranker-0.6B~\citep{li2026memreranker} & 23.84 & 56.01 & 66.76 & 77.00 & 70.32 & 68.80 & 65.00 & 82.21 & 57.06 \\
MemReranker-4B~\citep{li2026memreranker} & 23.72 & 56.62 & 68.35 & 75.50 & 70.71 & 70.00 & 66.31 & 81.69 & 57.69 \\
QRRanker~\citep{li2026query} & 19.49 & 50.98 & 62.75 & 63.00 & 62.41 & 62.42 & 58.83 & 73.59 & 51.76 \\
ConvMemory~\citep{pan2026convmemory,pan2026convmemoryv2} & 15.04 & 36.38 & 46.26 & 45.50 & 43.49 & 44.93 & 43.28 & 59.20 & 35.89 \\
\midrule
\textbf{ReMind} & \textbf{27.61} & \textbf{65.30} & \textbf{74.88} & \textbf{89.50} & \textbf{82.83} & \textbf{79.39} & \textbf{75.44} & \textbf{89.97} & \textbf{67.08} \\
\midrule
\multicolumn{10}{@{}l@{}}{\makebox[0pt][l]{\colorbox[gray]{0.95}{\makebox[\dimexpr\textwidth-2\fboxsep\relax]{\strut\textbf{\textit{Dense retriever}}}}}} \\
Base & 21.00 & 48.78 & 61.82 & 68.00 & 62.06 & 62.54 & 61.46 & 78.81 & 50.06 \\
\midrule
Qwen3-Reranker-0.6B~\citep{zhang2025qwen3} & 23.09 & 57.46 & 69.72 & 75.50 & 72.15 & 71.31 & 69.94 & 84.31 & 58.91 \\
Qwen3-Reranker-4B~\citep{zhang2025qwen3} & 25.72 & 66.43 & 83.01 & 85.50 & 83.80 & 83.90 & 82.68 & 91.02 & 69.03 \\
Qwen3-Reranker-8B~\citep{zhang2025qwen3} & 25.31 & 63.73 & 79.92 & 84.00 & 80.87 & 81.23 & 80.27 & 90.07 & 66.20 \\
KaLM-Nano~\citep{zhao2026kalm} & 27.18 & 63.63 & 79.37 & 88.00 & 81.33 & 81.40 & 80.29 & 91.80 & 65.75 \\
KaLM-Small~\citep{zhao2026kalm} & 27.82 & 67.21 & 82.41 & 90.50 & 85.01 & 84.68 & 83.42 & 93.63 & 69.37 \\
KaLM-Large~\citep{zhao2026kalm} & 27.72 & 67.53 & 83.21 & 90.00 & 85.52 & 85.34 & 83.93 & 93.88 & 69.74 \\
RankGPT~\citep{sun2023chatgpt} & 24.38 & 61.92 & 74.52 & 79.50 & 77.89 & 76.34 & 76.16 & 86.97 & 63.98 \\
\midrule
MemReranker-0.6B~\citep{li2026memreranker} & 25.66 & 63.19 & 78.95 & 84.50 & 80.06 & 80.35 & 79.36 & 90.33 & 65.29 \\
MemReranker-4B~\citep{li2026memreranker} & 25.23 & 63.76 & 80.46 & 81.50 & 79.78 & 80.94 & 80.28 & 88.73 & 65.80 \\
QRRanker~\citep{li2026query} & 21.54 & 55.73 & 72.86 & 70.50 & 69.41 & 72.03 & 70.95 & 81.06 & 57.16 \\
ConvMemory~\citep{pan2026convmemory,pan2026convmemoryv2} & 15.19 & 38.99 & 51.48 & 45.50 & 46.33 & 49.08 & 50.21 & 62.41 & 38.92 \\
\midrule
\textbf{ReMind-SFT} & \underline{28.97} & \underline{75.00} & \underline{92.43} & \underline{97.00} & \underline{95.74} & \underline{95.20} & \underline{93.62} & \underline{98.33} & \underline{78.68} \\
\textbf{ReMind} & \textbf{29.97} & \textbf{76.75} & \textbf{93.77} & \textbf{99.00} & \textbf{97.51} & \textbf{96.75} & \textbf{95.10} & \textbf{99.50} & \textbf{80.08} \\
\midrule
\multicolumn{10}{@{}l@{}}{\makebox[0pt][l]{\colorbox[gray]{0.95}{\makebox[\dimexpr\textwidth-2\fboxsep\relax]{\strut\textbf{\textit{RF-Mem}}}}}} \\
Base & 21.00 & 48.78 & 63.16 & 68.00 & 62.06 & 63.40 & 61.33 & 78.77 & 50.06 \\
\midrule
Qwen3-Reranker-0.6B~\citep{zhang2025qwen3} & 23.01 & 57.36 & 69.59 & 75.00 & 71.96 & 71.03 & 69.13 & 83.92 & 58.86 \\
Qwen3-Reranker-4B~\citep{zhang2025qwen3} & 25.62 & 66.28 & 82.09 & 85.00 & 83.56 & 83.19 & 81.66 & 90.76 & 68.87 \\
Qwen3-Reranker-8B~\citep{zhang2025qwen3} & 25.23 & 63.55 & 78.65 & 83.50 & 80.70 & 80.35 & 79.22 & 89.78 & 66.02 \\
KaLM-Nano~\citep{zhao2026kalm} & 27.08 & 63.24 & 78.67 & 87.50 & 80.79 & 80.78 & 79.24 & 91.38 & 65.31 \\
KaLM-Small~\citep{zhao2026kalm} & \underline{27.72} & 66.78 & 81.38 & \underline{90.00} & 84.32 & 83.76 & 82.23 & 93.21 & 68.85 \\
KaLM-Large~\citep{zhao2026kalm} & \underline{27.72} & \underline{67.46} & \underline{82.74} & \underline{90.00} & \underline{85.30} & \underline{84.94} & \underline{82.92} & \underline{93.79} & \underline{69.64} \\
RankGPT~\citep{sun2023chatgpt} & 24.53 & 62.61 & 74.87 & 80.00 & 78.59 & 76.77 & 76.21 & 87.34 & 64.66 \\
\midrule
MemReranker-0.6B~\citep{li2026memreranker} & 25.66 & 63.09 & 78.33 & 84.50 & 79.94 & 79.92 & 78.46 & 90.35 & 65.17 \\
MemReranker-4B~\citep{li2026memreranker} & 25.23 & 63.68 & 79.96 & 81.50 & 79.66 & 80.64 & 79.54 & 88.72 & 65.69 \\
QRRanker~\citep{li2026query} & 22.08 & 57.85 & 71.38 & 72.00 & 71.66 & 71.85 & 70.73 & 82.36 & 59.30 \\
ConvMemory~\citep{pan2026convmemory,pan2026convmemoryv2} & 15.32 & 39.55 & 52.17 & 46.00 & 47.03 & 49.70 & 50.25 & 62.88 & 39.60 \\
\midrule
\textbf{ReMind} & \textbf{29.72} & \textbf{76.52} & \textbf{92.74} & \textbf{98.50} & \textbf{97.11} & \textbf{95.89} & \textbf{93.73} & \textbf{99.25} & \textbf{79.82} \\
\bottomrule
\end{tabular*}
\end{table}

\section{Complete Results of the Sufficiency Predictor Analysis}
\label{app:sufficiency}
Table~\ref{tab:sufficiency} reports all downstream metrics for the comparison in Section~\ref{sec:sufficiency}.
\begin{table}[h]
\caption{Downstream question answering on PersonaBench with dense retrieval (\%). \#Mem is the average number of memories placed in the context. Bold marks the best result in each column among the fixed cutoffs and ReMind, and underline marks the second best; the two oracles are upper bounds and are not marked.}
\label{tab:sufficiency}
\small
\centering
\setlength{\tabcolsep}{4pt}
\renewcommand{\arraystretch}{1.0}
\begin{tabular*}{\textwidth}{@{\hspace{2pt}\extracolsep{\fill}}l*{5}{c}@{\hspace{2pt}}}
\toprule
\textbf{Method} & \textbf{Answer EM} & \textbf{Answer F1} & \textbf{Answer ROUGE-L} & \textbf{Answer LLM-judge} & \textbf{\#Mem} \\
\midrule
Fixed top-1 & 8.64 & 26.49 & 25.23 & 25.93 & 1.00 \\
Fixed top-2 & 11.11 & 28.87 & 29.95 & \underline{28.40} & 2.00 \\
Fixed top-3 & \underline{13.58} & \underline{30.88} & \underline{30.56} & \textbf{30.86} & 3.00 \\
Fixed top-4 & 9.88 & 24.83 & 23.32 & 25.93 & 4.00 \\
Fixed top-5 & 8.64 & 23.12 & 20.51 & 22.22 & 5.00 \\
\textbf{ReMind} & \textbf{14.81} & \textbf{31.91} & \textbf{31.24} & \textbf{30.86} & 3.81 \\
\midrule
Oracle stop & 16.05 & 33.41 & 34.17 & 32.10 & 2.12 \\
Gold memories & 16.05 & 34.79 & 34.81 & 33.33 & 2.02 \\
\bottomrule
\end{tabular*}
\end{table}

\paragraph{A matched memory budget.}

ReMind returns 3.81 memories per query on average, whereas the strongest fixed cutoff uses three memories, raising the question of whether the gain comes simply from using a larger memory budget. We therefore construct a query-independent randomized fixed policy with the same expected budget.
Among mixtures of the evaluated fixed cutoffs $K\in\{1,\ldots,5\}$, the strongest budget-matched policy assigns $K=3$ to $59.5\%$ of queries and $K=5$ to the remaining $40.5\%$, giving $0.595\times3+0.405\times5=3.81$ memories per query on average.

By linearity of expectation, this policy obtains expected scores of
11.58 exact match, 27.74 answer F1, 26.49 answer ROUGE-L, and 27.36 on the LLM judge. ReMind obtains 14.81, 31.91, 31.24, and 30.86, respectively, under the same average memory-count budget. Thus, matching the average number of retrieved memories does not account for the gain.

In this experiment, answer quality for the fixed-cutoff policy peaks at $K=3$ and decreases for larger cutoffs. Consequently, a query-independent policy cannot effectively use the additional budget because it does not know which queries require more evidence. ReMind instead allocates the number of memories on a per-query basis through its sufficiency predictor. Moreover, among the individually evaluated fixed cutoffs, none reaches the exact-match accuracy of ReMind: the best fixed cutoff obtains 13.58 EM, compared with 14.81 for ReMind.

\section{Inference Cost}
\label{app:inference-cost}
ReMind runs one ranking step for every memory it selects, so it takes more inference time than a single-pass ranker.
This section measures this time and compares it with the time that the reader needs to generate an answer.

\paragraph{Setting.}
We run ReMind on PersonaBench with the dense retriever for a fixed number of steps from one to five, on a single NVIDIA RTX 4090 GPU.
At each step, we measure the time of one forward pass that scores a single candidate given the query and the memories selected so far.
The candidates within a step are scored independently of each other, so they can be processed in parallel, and under full parallelism the time of a step equals the time of this forward pass.
The cumulative time adds up the step times and is therefore the ranking time of one query in this ideal setting.
For reference, we measure the average time that the reader of Section~\ref{sec:sufficiency}, Llama-3-8B-Instruct, takes to generate an answer on the same queries and the same GPU.
The ranking metrics are computed as in Section~\ref{sec:main-results}, where the positions after the last step are filled with the unselected candidates ordered by their scores from that step.

\begin{table}[h]
\caption{Inference cost analysis.}
\label{tab:inference-cost}
\small
\centering
\setlength{\tabcolsep}{8pt}
\renewcommand{\arraystretch}{1.0}
\begin{tabular}{lccc}
\toprule
\textbf{Steps} & \textbf{R@5 (\%)} & \textbf{N@5 (\%)} & \textbf{Cumulative time (ms)} \\
\midrule
1 & 76.28 & 72.17 & 28.93 \\
2 & 76.89 & 73.15 & 73.55 \\
3 & 77.51 & 73.75 & 132.81 \\
4 & 79.36 & 74.57 & 205.79 \\
5 & 79.92 & 75.01 & 297.48 \\
\midrule
Reader (answer generation) & -- & -- & 4829 \\
\bottomrule
\end{tabular}
\end{table}

\paragraph{Results.}
Table~\ref{tab:inference-cost} reports the results, and we can draw three conclusions.
First, the cumulative time grows faster than the number of steps, because every selected memory is added to the input and makes each later step longer.
As a result, five steps take about ten times as long as a single step.
Second, the ranking time is still small compared with answer generation: even with five steps, the cumulative time is 297.48 ms, which is about 6\% of the 4.83 s that the reader needs to generate an answer.
Since sufficiency stopping selects 3.81 memories per query on average (Table~\ref{tab:sufficiency}), the average ranking time in practice is lower than the five-step value.
Third, the additional steps keep improving the ranking: between the first and the fifth step, R@5 rises from 76.28 to 79.92 and NDCG@5 from 72.17 to 75.01.

\section{Complete Hyperparameter Analysis Results}
\label{app:complete-hyperparameter-results}
Table~\ref{tab:hyper-personabench} reports the complete hyperparameter analysis results on PersonaBench with dense retrieval, and its NDCG@3 column is plotted in Figure~\ref{fig:hyperparameter}.
\begin{table}[h]
\caption{Complete hyperparameter analysis results on PersonaBench with dense retrieval (\%). Bold marks the best result in each column within each hyperparameter and underline marks the second best.}
\label{tab:hyper-personabench}
\small
\centering
\setlength{\tabcolsep}{3pt}
\renewcommand{\arraystretch}{1.0}
\begin{tabular*}{\textwidth}{@{\hspace{2pt}\extracolsep{\fill}}l*{9}{c}@{\hspace{2pt}}}
\toprule
\textbf{Value} & \textbf{R@1} & \textbf{R@3} & \textbf{R@5} & \textbf{N@1} & \textbf{N@3} & \textbf{N@5} & \textbf{MAP} & \textbf{MRR} & \textbf{F1} \\
\midrule
\multicolumn{10}{@{}l@{}}{\makebox[0pt][l]{\colorbox[gray]{0.95}{\makebox[\dimexpr\textwidth-2\fboxsep\relax]{\strut\textbf{\textit{Negative ratio $\rho$}}}}}} \\
0.1 & 35.25 & 59.12 & 65.25 & 54.32 & 60.91 & 60.15 & 60.22 & 64.24 & 42.02 \\
0.2 & \textbf{45.66} & \textbf{66.21} & \textbf{77.24} & \textbf{69.14} & \textbf{70.42} & \textbf{71.84} & \textbf{69.55} & \textbf{76.89} & \textbf{45.37} \\
0.3 & \underline{40.64} & 63.97 & \underline{76.56} & \underline{60.49} & \underline{66.26} & \underline{68.43} & \underline{65.78} & \underline{71.63} & 43.87 \\
0.4 & 35.70 & 64.22 & 73.27 & 55.56 & 64.09 & 65.03 & 62.48 & 67.44 & \underline{44.18} \\
0.5 & 36.50 & \underline{65.35} & 75.39 & 55.56 & 64.20 & 65.73 & 63.06 & 68.73 & 43.60 \\
\multicolumn{10}{@{}l@{}}{\makebox[0pt][l]{\colorbox[gray]{0.95}{\makebox[\dimexpr\textwidth-2\fboxsep\relax]{\strut\textbf{\textit{Hard-negative count $K_n$}}}}}} \\
3 & \underline{45.66} & \underline{66.21} & \underline{77.24} & \underline{69.14} & \underline{70.42} & \underline{71.84} & \underline{69.55} & \underline{76.89} & \underline{45.37} \\
4 & \textbf{49.81} & \textbf{71.30} & \textbf{78.74} & \textbf{72.84} & \textbf{74.77} & \textbf{74.28} & \textbf{72.61} & \textbf{79.25} & \textbf{48.09} \\
5 & 37.92 & 64.14 & 71.03 & 56.79 & 63.99 & 63.60 & 61.61 & 68.43 & 43.15 \\
10 & 38.37 & 63.00 & 74.77 & 56.79 & 64.31 & 66.41 & 63.96 & 68.21 & 43.42 \\
\multicolumn{10}{@{}l@{}}{\makebox[0pt][l]{\colorbox[gray]{0.95}{\makebox[\dimexpr\textwidth-2\fboxsep\relax]{\strut\textbf{\textit{Positive orderings per query $P$}}}}}} \\
1 & 36.81 & 61.19 & 68.02 & 56.79 & 62.54 & 62.40 & 61.38 & 67.27 & 42.08 \\
3 & \textbf{45.82} & \textbf{73.99} & \textbf{82.12} & \underline{66.67} & \textbf{73.98} & \textbf{74.36} & \textbf{71.21} & \underline{75.98} & \textbf{49.47} \\
5 & \underline{45.66} & \underline{66.21} & \underline{77.24} & \textbf{69.14} & \underline{70.42} & \underline{71.84} & \underline{69.55} & \textbf{76.89} & \underline{45.37} \\
7 & 35.16 & 65.00 & 71.46 & 55.56 & 64.38 & 63.65 & 61.82 & 67.38 & 44.78 \\
9 & 38.17 & 63.19 & 71.21 & 58.02 & 63.87 & 64.24 & 62.67 & 68.97 & 42.72 \\
\multicolumn{10}{@{}l@{}}{\makebox[0pt][l]{\colorbox[gray]{0.95}{\makebox[\dimexpr\textwidth-2\fboxsep\relax]{\strut\textbf{\textit{Sufficiency threshold $\tau$}}}}}} \\
0.2 & \textbf{45.66} & \textbf{66.79} & \textbf{78.48} & \textbf{69.14} & \textbf{70.46} & \textbf{72.47} & \textbf{69.88} & \textbf{77.43} & \underline{45.26} \\
0.4 & \textbf{45.66} & 65.80 & \underline{78.23} & \textbf{69.14} & 70.13 & \underline{72.17} & \underline{69.73} & \underline{77.27} & 44.96 \\
0.5 & \textbf{45.66} & \underline{66.21} & 77.24 & \textbf{69.14} & \underline{70.42} & 71.84 & 69.55 & 76.89 & \textbf{45.37} \\
0.6 & \textbf{45.66} & \underline{66.21} & 77.24 & \textbf{69.14} & \underline{70.42} & 71.84 & 69.58 & 76.88 & \textbf{45.37} \\
0.8 & \textbf{45.66} & \underline{66.21} & 77.24 & \textbf{69.14} & \underline{70.42} & 71.86 & 69.63 & 76.88 & \textbf{45.37} \\
\multicolumn{10}{@{}l@{}}{\makebox[0pt][l]{\colorbox[gray]{0.95}{\makebox[\dimexpr\textwidth-2\fboxsep\relax]{\strut\textbf{\textit{Joint-loss weight $\lambda$}}}}}} \\
0.2 & 34.57 & 56.32 & 69.88 & 54.32 & 58.25 & 61.24 & 58.43 & 64.42 & 39.65 \\
0.4 & 39.44 & \underline{62.92} & 68.87 & \underline{60.49} & \underline{65.01} & 64.31 & 63.53 & 69.64 & \underline{43.31} \\
0.6 & \textbf{45.66} & \textbf{66.21} & \textbf{77.24} & \textbf{69.14} & \textbf{70.42} & \textbf{71.84} & \textbf{69.55} & \textbf{76.89} & \textbf{45.37} \\
0.8 & 38.37 & 62.65 & \underline{75.12} & 56.79 & 63.91 & 65.94 & 63.11 & 68.32 & 43.15 \\
1 & \underline{40.27} & 61.81 & 74.67 & \underline{60.49} & 63.68 & \underline{66.57} & \underline{64.50} & \underline{69.77} & 41.85 \\
\bottomrule
\end{tabular*}
\end{table}

\section{Complete Results by Query Complexity}
\label{app:query-complexity}
Table~\ref{tab:query-complexity-complete} adds Recall@3 to the comparison in Section~\ref{sec:query-complexity}.
\begin{table}[h]
\caption{NDCG@3 and Recall@3 by the number of gold memories a query needs, on PersonaBench with dense retrieval (\%). Bold marks the best result in each column and underline marks the second best.}
\label{tab:query-complexity-complete}
\small
\centering
\setlength{\tabcolsep}{4pt}
\renewcommand{\arraystretch}{1.0}
\begin{tabular*}{\textwidth}{@{\hspace{2pt}\extracolsep{\fill}}l*{6}{c}@{\hspace{2pt}}}
\toprule
& \multicolumn{2}{c}{\textbf{2 gold memories}} & \multicolumn{2}{c}{\textbf{3 gold memories}} & \multicolumn{2}{c}{\textbf{4+ gold memories}} \\
\cmidrule(lr){2-3}\cmidrule(lr){4-5}\cmidrule(lr){6-7}
\textbf{Method} & \textbf{N@3} & \textbf{R@3} & \textbf{N@3} & \textbf{R@3} & \textbf{N@3} & \textbf{R@3} \\
\midrule
MemReranker-4B & 56.45 & \underline{58.33} & \underline{66.67} & \underline{61.11} & 63.75 & 38.86 \\
ReMind-SFT & \underline{60.22} & \underline{58.33} & 62.76 & \underline{61.11} & \underline{69.07} & \underline{44.65} \\
\textbf{ReMind} & \textbf{63.99} & \textbf{66.67} & \textbf{69.55} & \textbf{66.67} & \textbf{81.74} & \textbf{51.32} \\
\bottomrule
\end{tabular*}
\end{table}

\end{document}